\documentstyle[11pt]{article}
\title{Viscosity of Colloidal Suspensions}
\author{R. Verberg, I. M. de Schepper\\
I. R. I., Delft University of Technology\\
2629 JB Delft, The Netherlands\\
and\\
E. G. D. Cohen\\
The Rockefeller University\\
New York, NY 10021, USA} 
\date{  }
\begin{document}
\input{psfig}
\maketitle

\begin{abstract}
Simple expressions are given for the Newtonian viscosity $\eta_N(\phi)$
as well as the viscoelastic behavior of the viscosity $\eta(\phi,\omega)$ 
of neutral monodisperse hard sphere colloidal suspensions as a function of 
volume fraction $\phi$ and frequency $\omega$ over the entire fluid range,
i.e., for volume fractions $0 < \phi < 0.55$.  These expressions are 
based on an approximate theory which considers the viscosity as composed 
as the sum of two relevant physical processes:
$\eta (\phi,\omega) = \eta_{\infty}(\phi) + \eta_{cd}(\phi,\omega)$, where 
$\eta_{\infty}(\phi) = \eta_0 \chi(\phi)$ is the infinite frequency (or very
short time) viscosity, with $\eta_0$ the solvent viscosity, $\chi(\phi)$ the 
equilibrium hard sphere radial 
distribution function at contact, and $\eta_{cd}(\phi,\omega)$
the contribution due to the diffusion of the colloidal particles out of 
cages formed by their neighbors, on the P\'{e}clet time scale $\tau_P$,
the dominant physical process in concentrated colloidal suspensions.
The Newtonian viscosity $\eta_N(\phi) = \eta(\phi,\omega = 0)$ agrees very 
well with the extensive experiments of Van der Werff et al and others.
Also, the asymptotic behavior for large $\omega$ is of 
the form $\eta_{\infty}(\phi) + A(\phi)(\omega \tau_P)^{-1/2}$, in agreement
with these experiments, but the theoretical coefficient $A(\phi)$ differs by a
constant factor $2/\chi(\phi)$ from the exact coefficient, computed from the
Green-Kubo formula for $\eta(\phi,\omega)$.  This still enables us to predict 
for practical purposes the visco-elastic behavior of monodisperse spherical
colloidal suspensions for all volume fractions by a simple time rescaling. 
\end{abstract}
\newpage
\hsize=6in
\hoffset=-.5in
\baselineskip=1.5\baselineskip
\section{Introduction}
In a number of previous papers we have discussed the Newtonian viscosity
as well as the visco-elastic behavior of concentrated colloidal suspensions,
consisting of monodisperse neutral hard sphere particles$^{[1-4]}$.  The 
motivation was to understand theoretically the very extensive viscosity
measurements on colloidal suspensions carried out by Van der Werff 
et al in Utrecht$^{[5,6]}$.  In particular, these experiments on carefully
prepared systems seemed to be an ideal testing ground for the theory.
In this paper a more complete and detailed account of the viscous 
behavior of colloidal suspensions over their fluid range will be given.

Our theoretical approach is based on two physical processes related to
the two widely separated basic time scales in a colloidal suspension: the
Brownian time $\tau_B \sim 10^{-8}$s, during which a single Brownian particle 
forgets its initial velocity and the interaction time or P\'{e}clet time 
$\tau_P = \sigma^2/4 D_0 \sim 10^{-3}$s, during and beyond which Brownian 
particle interactions take
place.  Here $\sigma$ is the diameter of the hard sphere colloidal particles
and $D_0$ the Stokes-Einstein colloidal particle diffusion coefficient at
infinite dilution.  The viscosity is consequently considered as composed of 
a sum of contributions which take place on a short and a long 
time scale.  Although the theory is constructed for concentrated 
colloidal suspensions with volume fractions $0.3 < \phi < 0.55$, it
appears that the theory also gives good numerical results for lower
concentrations, so that effectively formulae are obtained which cover
the entire fluid range $0 < \phi < 0.55$.  Here $\phi = n \pi \sigma^3/6$,
where $n$ is the number density of the hard sphere colloidal particles.

The suspension is considered as a homogeneous fluid consisting of spherical
particles immersed in a continuum solvent.  As a consequence formulae 
derived for simple homogeneous fluids in general - like the Irving-Kirkwood
expression for the pressure tensor$^{[7,8]}$ or the Green-Kubo formula for the
viscosity$^{[9]}$ - are also assumed to be applicable here.  The formulae
for the viscous behavior are derived under a number of assumptions, which we
will try to justify physically as well as possible, but which, considering the 
complexity of this strongly interacting system, we have not been able to 
derive from first principles or justify completely.

The two basic physical processes we referred to are 

1. at short times $t \leq \tau_B \ll \tau_P$ and nonzero 
concentrations, the viscosity of the suspension effectively increases when 
compared to that of the (pure) solvent viscosity $\eta_0$ at infinite dilution,
due to the finite probability to find two particles at contact; 

2. at long times $t \sim \tau_P \gg \tau_B$, the difficulty of a 
Brownian particle to diffuse out of the cage formed around it by its neighbors,
characterized by a cage-diffusion coefficient $D_c(k;\phi)$.  

As to 1., the probability to find two particles in the suspension 
at contact is given by the equilibrium radial distribution function at contact:
$g_{eq}(\sigma; \phi) \equiv \chi(\phi)^{[10]}$, which follows from the canonical
distribution of the hard sphere colloidal particles.  As a result, the 
effective very high frequency viscosity of the suspension satisfies 
$\eta_{\infty}(\phi) = \eta_0 \chi(\phi)$, a relation which is consistent 
with experiment over the entire fluid range$^{[4]}$.  Similarly, the short time
self-diffusion coefficient
of the Brownian particles past each other is decreased from the 
Stokes-Einstein value $D_0$ at infinite dilution, to a value 
$D_s(\phi) = D_0/\chi(\phi)$, since $\chi(\phi)$ also gives the increase
in the binary collision frequency in a dense hard sphere gas in equilibrium as 
compared to that in a dilute gas.  Also this relation has been confirmed 
by experiment$^{[4]}$.                         

As to 2., the cage diffusion coefficient $D_c(k;\phi)$ refers to the 
diffusion of a particle out of a cage formed by its neighbors when the 
particles are distributed periodically in the solvent with a wave number $k$.
For concentrated suspensions one should bear in mind that a typical wave
number is $k \approx k^* = 2 \pi/\sigma$, corresponding to a surface to 
surface distance of two neighboring Brownian particles of typically 1/10 of 
their diameter $\sigma$, so that the particles ``rattle'' in their cages
before they diffuse out in a time of the order of $\tau_P \approx
\tau_c(k^*;\phi) = 1/D_c(k^*;\phi)k^{*^{2}}$. In fig.1 $\tau_c(k;\phi)/
\tau_P$ is plotted as a function of $\kappa = k \sigma$ for four 
values of $\phi$. $\tau_c(k;\phi)$ and $\tau_P$ are clearly of the same
order of magnitude, the pronounced maximum of $\tau_c(k;\phi)$ at $k = k^*$ 
corresponding to the ``rattling in the cage''.   An explicit expression 
for the cage diffusion coefficient $D_c(k;\phi)$ has been obtained from kinetic
theory$^{[11]}$. Since $D_c(k;\phi)$ also characterizes the decay 
of a spontaneous density fluctuation of wave number $k$ in the 
suspension$^{[12]}$, it can be measured by light or neutron scattering
and the expression we give for it below has been shown to be in good agreement
with such experiments$^{[13]}$.

To incorporate the cage-diffusion process, i.e., $D_c(k;\phi)$ into the 
theory, we need to go to a Fourier (i.e., ${\bf{k}}$-) representation,
while the starting point of our theory, the two particle Smoluchowski
equation$^{[14]}$, is expressed in ordinary (i.e., ${\bf{r}}$-) space.  This will
introduce a fundamental difficulty in the development of the theory, since
the impenetrability of two hard sphere particles, which is easily accounted 
for in ${\bf{r}}$- space, will be violated in our theory in ${\bf{k}}$-space,
a point that will be discussed further below.

The paper is constructed as follows.  In section 2 we give the basic 
equations for the viscosity of the colloidal suspension and for the
nonequilibrium pair distribution function of the colloidal particles to 
obtain this viscosity from a solution of the latter equation.  In section 3 
this solution is used to obtain an explicit expression for the visco-elastic 
behavior $\eta(\phi,\omega)$ of the suspension.  Section 4 gives a simple 
formula for the zero-frequency or Newtonian viscosity $\eta_N(\phi) = 
\eta(\phi,\omega = 0)$, while section 5 contains the 
visco-elastic behavior of the fluid for finite frequencies.  In section 6
the approach of $\eta(\phi,\omega)$ to its asymptotic value 
$\eta_{\infty}(\phi)$, via a behavior $\sim A(\phi)(\omega \tau_P)^{-1/2}$, is 
discussed and exact results for the coefficient $A(\phi)$ are compared with our 
theory and with experiment.  In section
7 the behavior of $\eta(\phi,\omega)$ for small $\omega$ is given and 
section 8 discusses a number of issues raised by the results obtained in the
paper, especially in connection with the good agreement with experiment,
in spite of the apparent neglect of hydrodynamic interactions between the 
Brownian particles.\\  
\section{Basic Equations}

The shear viscosity we are concerned with in this paper is defined as the
linear response of the suspension to an applied shear rate $\gamma (t) =
\gamma_0 e^{-i \omega t}$ with finite frequency $\omega$ and vanishing amplitude
$\gamma_0$, or equivalently by
\begin{equation}
P_{xy}(\phi,\omega,\gamma_0,t) = - \eta(\phi,\omega,\gamma_0,t) \gamma(t)
\end{equation}
Here $P_{xy}$ is the $xy$-component of the pressure tensor of the
suspension, defined by
\begin{equation}
P_{xy}(\phi,\omega,\gamma_0,t) = P_{xy,s}(\phi,\gamma_0,t) + P_{xy,d}
(\phi,\omega,\gamma_0,t)
\end{equation}
where $P_{xy,s}(\phi,\gamma_0,t)$ is the static contribution ($\omega = \infty$)
to the $xy$-component of the pressure tensor and
$P_{xy,d}(\phi,\omega,\gamma_0,t)$ the dynamic contribution given by$^{[7]}$
\begin{equation}
P_{xy,d}(\phi,\omega,\gamma_0,t) =
-\frac{1}{2V} < \sum^N_{j \neq i = 1} r_{ij,x}
\frac{\partial V(r_{ij})}{\partial r_{i,y}} >_{n.e.}
\end{equation}
Here $V$ is the volume of the system, ${\bf{r}}_i$ the position of particle 
$i \: (i=1,...,N),{\bf{r}}_{ij} = {\bf{r}}_i - {\bf{r}}_j, V(r_{ij})$ the
interparticle potential between particles $i$ and $j$ at a distance
$r_{ij} = |{\bf{r}}_{ij}|$ and the non-equilibrium average $< \: >_{n.e.}$ is
taken with respect to a nonequilibrium distribution function derived from the
$N$-particle
Smoluchowski equation for a suspension under shear rate $\gamma(t)$.
Kinetic contributions to the pressure tensor are not considered in such a 
description of the system.

The static contribution follows from the limit $\omega \rightarrow \infty$ when
the dynamic contribution to the pressure tensor becomes zero, leaving in
eq.(1) only
\begin{equation}
P_{xy}(\phi,\omega=\infty,\gamma_0,t) = P_{xy,s}(\phi,\gamma_0,t) = 
- \eta_{\infty}(\phi) \gamma(t).
\end{equation}
Carrying out the implied integration on the right hand side (r.h.s.) of eq.(3)
over the positions of all $(N-2)$ particles,
but the particles 1 and 2, introducing center of mass and relative 
coordinates by ${\bf{R}} = ({\bf{r}}_1 + {\bf{r}}_2)/2$ and ${\bf{r}} =
{\bf{r}}_1 -{\bf{r}}_2$, respectively, and carrying out the integration over 
${\bf{R}}$, one obtains for the dynamic contribution to the pressure tensor
\begin{equation}
P_{xy,d} (\phi,\omega,\gamma_0,t) = -\frac{n^2}{2} \int d{\bf{r}}
g({\bf{r}};\phi,\omega,\gamma_0,t) x \frac{\partial V(r)}{\partial y}
\end{equation}
This gives with eqs.(2) and (4) the following expression for the total
pressure tensor:
\begin{equation}
P_{xy} (\phi,\omega,\gamma_0,t) =
-\eta_{\infty}(\phi) \gamma(t) -\frac{n^2}{2} \int d{\bf{r}}
g({\bf{r}};\phi,\omega,\gamma_0,t) x \frac{\partial V(r)}{\partial y}
\end{equation}
Here $n^2g({\bf{r}}; \phi,\omega,\gamma_0,t)$ is the nonequilibrium pair
distribution function, giving the average number of colloidal particle
pairs at a separation ${\bf{r}}$ in the suspension at a number density
$n$ of the colloidal particles, so that $g({\bf{r}}; \phi,\omega,\gamma_0,t)$
is the nonequilibrium generalization of the radial distribution function
$g_{eq}(r;\phi)$ in equilibrium, when $\gamma_0 = 0$.  Introducing then:
\newcounter{eqnnum}
\newcounter{eqnletter}
\renewcommand{\theequation}{\arabic{eqnnum}\alph{eqnletter}}
\setcounter{eqnnum}{7}
\setcounter{eqnletter}{1}
\begin{equation} 
g({\bf{r}};\phi,\omega,\gamma_0,t) = g_{eq}(r;\phi) + \delta g 
({\bf{r}};\phi,\omega,\gamma_0) e^{-i \omega t} 
\end{equation}
we have for $\gamma_0 \rightarrow 0$
\addtocounter{eqnletter}{1}
\begin{equation}
\delta g({\bf{r}};\phi,\omega,\gamma_0) = \gamma_0 \delta g({\bf{r}};
\phi,\omega) + O(\gamma_0^2)
\end{equation}
and one finds from eq.(6) that in the limit of vanishing shear rate 
$\gamma_0 \rightarrow 0, P_{xy}(\phi,\omega,\gamma_0,t)$ is proportional
to $\gamma(t)$ since the contribution of $g_{eq}(r;\phi)$ vanishes.  Then
in eq.(1), the viscosity $\eta(\phi,\omega) =
{\rm{lim}}_{\gamma_{0}\rightarrow 0} \eta(\phi,\omega,\gamma_0,t)$
is independent of $\gamma_0$ and $t$ and given by:
\renewcommand{\theequation}{\arabic{eqnnum}}
\addtocounter{eqnnum}{1}
\begin{equation}
\eta(\phi,\omega) = \eta_{\infty}(\phi) + \frac{1}{2} n^2 \int d{\bf{r}} 
\delta g ({\bf{r}}; \phi,\omega) x \frac{\partial V(r)}{\partial y}
\end{equation}
  
An approximate equation for $\delta g({\bf{r}};\phi,\omega)$ can be 
obtained in the following way.  Neglecting the hydrodynamical interactions
between the Brownian particles transmitted via the solvent, the $N$-particle
Smoluchowski equation for this case in a shear field $\gamma(t)$ can be
integrated over the positions of all ($N - 2$) particles but the two particles
1 and 2.  This leads to an equation for the nonequilibrium pair distribution
function, involving the nonequilibrium three particle distribution function.
Neglecting the latter, i.e. restricting ourselves to low densities
(to $O(\phi^2)$), transforming to
center of mass and relative coordinates
of the two particles 1 and 2, neglecting the dependence on the former, 
i.e.,  assuming spatial homogeneity and using $g_{eq}(r;\phi) =
\exp(-\beta V(r))$, one obtains the following equation for 
$g({\bf{r}};\phi,\omega,\gamma_0,t)$:
\addtocounter{eqnnum}{1}
\begin{equation}
[\frac{\partial}{\partial t} + 2 \beta D_0 {\bf{\nabla}} \cdot {\bf{F}}
({\bf{r}}) - 2 D_0 \nabla^2 + \gamma(t) x \frac{\partial}{\partial y}]
\:  g({\bf{r}};\phi,\omega,\gamma_0,t)=0
\end{equation}
Here ${\bf{F}}({\bf{r}}) = - \nabla V(r)$ is the force on particle 1 
at a separation ${\bf{r}}$ from particle 2, $\beta = 1/k_BT$, with $k_B$ 
Boltzmann's
constant and $T$ the absolute temperature.  Eq.(9) has been considered for 
charged colloidal suspensions in the stationary state, i.e. for $\omega = 0$
by Dhont et al$^{[15]}$.

With eq.(7), eq.(9) can be written as an equation for $\delta g({\bf{r}};
\phi,\omega)$:
\addtocounter{eqnnum}{1}
\begin{equation}
[-i\omega + 2 \beta D_0 \nabla \cdot {\bf{F}}({\bf{r}}) - 2 D_0 \nabla^2]
\delta g({\bf{r}};\phi,\omega) = - x\frac{\partial}{\partial y} 
e^{-\beta V(r)}
\end{equation}
which has been solved exactly by Cichocki and Felderhof$^{[16]}$ for hard
sphere particles (cf.Appendix A).

From now on we shall explicitly use a hard sphere potential unless
specified otherwise.  Neglecting then
the force term on the left hand side (l.h.s.) of eq.(10) and taking
the Fourier transform of eq.(10) with respect to ${\bf{r}}$, an equation is
obtained for:
\renewcommand{\theequation}{\arabic{eqnnum}\alph{eqnletter}}
\setcounter{eqnnum}{11}
\setcounter{eqnletter}{1}
\begin{equation}
\delta S({\bf{k}};\phi,\omega) = n \int d{\bf{r}} e^{i{\bf{k}} \cdot {\bf{r}}}
\delta g({\bf{r}};\phi,\omega)
\end{equation}
Using that
\addtocounter{eqnletter}{1}
\begin{equation}
S_{eq}(k;\phi) = 1+n \int d{\bf{r}} e^{i{\bf{k}} \cdot {\bf{r}}}
[g_{eq}(r;\phi)-1]
\end{equation}
is the static structure factor in equilibrium in general, the equation for 
$\delta S({\bf{k}}; \phi,\omega)$ derived from eq.(10) becomes:
\renewcommand{\theequation}{\arabic{eqnnum}}
\addtocounter{eqnnum}{1}
\begin{equation}
[-i\omega + 2 D_0 k^2] \delta S(k;\phi,\omega) = 24 \phi \frac{k_x k_y}{k^2}
j_2(k \sigma)
\end{equation}
where $j_2(k\sigma)$ is the spherical Bessel function of order 2$^{[17]}$.

As pointed out in the Introduction, the neglect of the force term (which is
only
justified for $r > \sigma$) in taking the Fourier transform of eq.(10), is the 
source of an error in the theory used in this paper to obtain the viscosity 
$\eta(\phi,\omega)$. A more detailed discussion of the nature of this error,
its consequences and a way to partially correct for it can be found in
section 6 and Appendix A.

Eq.(12) is only valid for dilute suspensions where $g_{eq} (r;\phi) = \exp
(-\beta V(r))$, i.e. $S_{eq}(k;\phi) = 1 - 24 \phi j_1 (k\sigma)
/ (k\sigma)$, and where the basic
diffusion process of the two particles is free diffusion, represented by the
term $2 D_0 k^2$ on the l.h.s. of eq.(12).

In order to
obtain an equation for concentrated colloidal suspensions we make two
corrections, a static one and a dynamic one. The first one
replaces the low density expression for the equilibrium static structure 
factor, used to derive eq.(12) from eqs.(10) and (11), by the complete 
$S_{eq}(k;\phi)$ for concentrated colloidal suspensions.
For the second correction we postulate that 
for such suspensions the  basic diffusion process is cage diffusion rather 
than free diffusion.  An expression for the relaxation time $\tau_c(k;\phi)$
for cage diffusion for concentrated colloidal suspensions has been derived
before from the kinetic theory of a dense
fluid of hard spheres, as the (scaled) reciprocal of the lowest eigenvalue 
$D_c(k;\phi)k^2$ of a linear generalized kinetic operator, discussed 
elsewhere$^{[11-13]}$: 
\addtocounter{eqnnum}{1}
\begin{equation}
\frac{1}{\tau_c(k;\phi)} = D_c (k,\phi)k^2 = \frac{D_0 k^2}{\chi (\phi)S_{eq}
(k;\phi)} d(k) 
\end{equation}
Here $D_c(k;\phi)$ is the cage diffusion coefficient,  $S_{eq}(k;\phi)$ is
again the equilibrium static structure for all $\phi$ and
$d(k) = 1/(1 - j_0(k) + 2j_2(k))$ a combination of spherical Bessel 
functions $j_{\ell} (k)$ of order $\ell = 0$ and $\ell = 2^{[17]}$.  Writing
\addtocounter{eqnnum}{1}
\begin{equation}
\frac{1}{\tau_c(k;\phi)} = \omega_H(k;\phi)
\end{equation}
the frequency $\omega_H(k;\phi)$ is the half-width at half height of the
dynamical structure factor $S_{eq}(k;\omega)$ of the suspension in 
equilibrium, which is the quantity that can be measured in light scattering
experiments.  The equality (14) is very well supported by 
experiment$^{[13]}$. Then eq.(12), becomes with eqs.(13) and (14):
\addtocounter{eqnnum}{1}
\begin{equation}
[-i\omega + 2 \omega_H(k;\phi)] \delta S \: ({\bf{k}};\phi,\omega) = 
k_y \frac{\partial}{\partial k_x} \; S_{eq}(k;\phi)
\end{equation}
which has the solution:
\addtocounter{eqnnum}{1}
\begin{equation}
\delta S({\bf{k}};\phi,\omega) = \frac{k_xk_y}{k}
\frac{S'_{eq}(k;\phi)}{2 \omega_H(k;\phi) - i \omega}
\end{equation}
where $S'_{eq}(k;\phi) = d S_{eq}(k;\phi)/d k$.

We note that $S_{eq}(k;\phi)$ has a very sharp maximum at 
$k \sim k^* = 2\pi/\sigma$ at high densities$^{[13]}$ indicating a quasi 
periodic ordering of the colloidal particles on the length scale $\sigma$ in 
cages.

Eq.(16) for $\delta S({\bf{k}}; \phi,\omega)$ can be used to
compute $\eta(\phi,\omega)$ with eqs.(8) and (11).  This will be shown in 
the next section.
\section{General expression for the viscosity}

In order to use eq.(16) for $\delta S({\bf{k}}; \phi, \omega)$ to compute 
$\eta (\phi, \omega)$ we must Fourier transform eq.(8).  For a hard sphere 
potential  such a transformation is not possible.  Therefore we replace 
in the spirit of the mean spherical approximation$^{[18]}$, $V(r)$ on the
r.h.s. of eq.(8) by the equilibrium hard sphere direct correlation 
function $C_{eq}(r; \phi)$, i.e.,
\addtocounter{eqnnum}{1}
\begin{equation}
V(r) \rightarrow -k_BT C_{eq}(r;\phi)
\end{equation}
As discussed in Section 6 and Appendix A, this replacement corrects partially
for the neglect of the force term on the l.h.s. of eq.(10), which leads to 
unphysical contributions from overlapping particle configurations.
Fourier transforming then eq.(8) by using Parcival's theorem on the r.h.s.
and that the Fourier transform $C_{eq}(k;\phi)$ of $C_{eq}(r;\phi)$ is related
to $S_{eq}(k;\phi)$ by:
\addtocounter{eqnnum}{1}
\begin{equation}
n C_{eq}(k;\phi) = 1- \frac{1}{S_{eq}(k;\phi)}
\end{equation}
one obtains straightforwardly from eqs.(8) and (11) the expression:
\addtocounter{eqnnum}{1}
\begin{equation}
\eta(\phi,\omega) = \eta_{\infty}(\phi) + \frac{k_BT}{1 6\pi^3}
\int d{\bf{k}} \frac{k_xk_y}{k} \frac{S'_{eq}(k;\phi)}{S_{eq}(k;\phi)^2} 
\delta S({\bf{k}};\phi,\omega)
\end{equation}
Substituting eq.(16) into eq.(19) we obtain, after an angular integration 
in ${\bf{k}}$-space:
\addtocounter{eqnnum}{1}
\begin{equation}
\eta(\phi,\omega) = \eta_{\infty}(\phi) + \frac{k_BT}{60\pi^2} \int^{\infty}_0
dk k^4 [\frac{S'_{eq}(k;\phi)}{S_{eq}(k;\phi)}]^2 \frac{1}{2\omega_H(k;\phi)
-i\omega}
\end{equation}
for the visco-elastic behavior of the suspension.

In so far as the integrand in the second term on the r.h.s. of eq.(20) 
contains the eigenvalues $(\omega_H(k;\phi))$ and amplitudes 
$(S'_{eq}(k;\phi)/S_{eq}(k;\phi))$ of {\bf{two}} cage diffusion modes, 
this term can be called a mode-mode 
coupling contribution to the viscosity.  The difference with the usual 
mode-mode coupling contributions is that here two cage-diffusion modes, which 
describe the diffusion process in and out of two neighboring particles' cages,
rather than two hydrodynamic modes (as occur in the long time tails 
or vortex diffusion$^{[19]}$) are used.  We also note that the same expression 
(20) for $\eta(\phi,\omega)$ can be derived for $\omega = 0$,
by a direct application of mode-mode coupling theory to the Green-Kubo
expression for $\eta (\phi, \omega = 0)^{[20]}$.  Since the complete derivation 
appears not to be in the literature, we briefly sketch it in Appendix B. 
For the concentrated suspensions we are mainly interested in here, the most 
important contributions to the integral in eq.(20) come from values of 
$k \approx k^*$. 

We note that the $k$-integral on the r.h.s. of eq.(20) is convergent for
all $\omega$, since the integrand vanishes for $k \rightarrow 0$ and the
asymptotic behavior for $k \rightarrow \infty$ is $\sim k^{-2}$,
as for large $k$:
\renewcommand{\theequation}{\arabic{eqnnum}\alph{eqnletter}}
\setcounter{eqnnum}{21}
\setcounter{eqnletter}{1}
\begin{equation}
S_{eq}(k;\phi)=1-24 \phi \chi(\phi) \frac{j_1(k\sigma)}{k\sigma} [1+O(k^{-2})];
\end{equation}
\addtocounter{eqnletter}{1}
\begin{equation}
S'_{eq}(k;\phi) = 24 \phi \chi(\phi) \frac{j_2(k\sigma)}{k} [1+O(k^{-2})];
\end{equation}
\addtocounter{eqnletter}{1}
\begin{equation}
\omega_H(k;\phi) = \frac{D_0}{\chi(\phi)} k^2[1 + O(k^{-2})]
\end{equation}
This implies that the second term on the r.h.s. of eq.(20) vanishes for 
$\omega \rightarrow \infty$, as it should, since $\eta(\phi, \infty) \equiv 
\eta_{\infty} (\phi)$ by definition.

We still have to obtain $\eta_{\infty}(\phi)$, in order to compute
$\eta(\phi,\omega)$.  One often writes $\eta_{\infty} (\phi) = \eta_0^{[21]}$,
i.e., equates $\eta_{\infty}(\phi)$ to the pure solvent viscosity, but this 
seems only correct
for dilute solutions.  For concentrated solutions, we propose to set:
\renewcommand{\theequation}{\arabic{eqnnum}}
\addtocounter{eqnnum}{1}
\begin{equation}
\eta_{\infty}(\phi) = \eta_0 \chi(\phi)
\end{equation}
implying that the effective viscosity of the suspension at very high 
frequencies 
is not only determined by the solvent viscosity but increased by the
fraction of colloidal particle pairs at contact, $\chi(\phi)$.
These touching, i.e. colliding particles, increase the effective  
viscosity proportional to the number of such pairs present in
the suspension, because they increase the viscous dissipation in the 
suspension due
to the instantaneous exchange of momentum during their collisions,
no matter how short the time scale.  They constitute therefore
an instantaneous contribution to $\eta(\phi, \omega)$.  Since$^{[10]}$
\addtocounter{eqnnum}{1}
\begin{equation}
\chi(\phi) = 1 + \frac{5}{2} \phi + 4.59 \phi^2 + O(\phi^3)
\end{equation}
eq.(22) reduces to the usual expression for $\eta_{\infty}(\phi)$ at small 
concentrations (see also section 8, sub.3).

In Fig.2 the behavior of $\eta_{\infty}(\phi)/\eta_0 = \chi(\phi)$ is compared
with the reduced viscosity measurements by Van der Werff et al$^{[5]}$ and Zhu
et al$^{[22]}$ at very high frequencies for $\phi$ over the entire fluid range
$0 < \phi < 0.55$.
Here we used the Carnahan-Starling approximation$^{[10]}$ 
\addtocounter{eqnnum}{1}
\begin{equation}
\chi (\phi) = \frac{1-0.5\phi}{(1-\phi)^3}
\end{equation}
which is very accurate for all such $\phi$.  The agreement between theory
and experiment is good, thus confirming eq.(22).  We note, however, that a 
theoretical derivation of eq.(22) is lacking (see section 8, sub 3).

We also included in fig.2 the values for $\eta_\infty(\phi)$ as obtained by
Cichocki and Felderhof$^{[23]}$.  These values differ slightly from those 
used by Van der Werff et al, since they obtained $\eta_\infty(\phi)$ by
fitting the tails of the data for large $\omega$ to $\eta_{\infty}(\phi) +
A(\phi)\sqrt{\omega \tau_P}$, instead of using a fit for all $\omega$.  We
used Cichocki and Felderhof's values for $\eta_\infty(\phi)$ throughout
the paper (cf.table 2).

We remark that eq.(20) with eq.(22) and all the equations following from them,
like eq.(25) in the next session, contain no adjustable parameters and are
comletely determined by those charaterising the system: the viscosity of
the solvent $\eta_0$, the volume fraction $\phi$ (or equivalantly the number
density $n$) and the diameter $\sigma$ of the colloidal particles.

In the next two sections, we will compare the concentration dependence
of the eq.(20) for the Newtonian viscosity $\eta_N(\phi) = \eta (\phi,\omega = 
0)$ and the concentration and frequency dependency of $\eta (\phi, \omega)$ of
eq.(20) with the experimental results of Van der Werff et al and others, 
respectively.
\section{Newtonian viscosity}

Setting $\omega = 0$ in eq.(20) and using eqs.(13), (14) and (22), we obtain 
the following simple expression for the Newtonian viscosity:
\addtocounter{eqnnum}{1} 
\begin{equation}
\eta_N(\phi) = \eta_0 \chi (\phi) [1 + \frac{1}{40\pi} \int^{\infty}_0
d \kappa \kappa^2 \frac{[S'_{eq}(\kappa;\phi)]^2}{S_{eq}(\kappa;\phi)d(\kappa)}]
\end{equation}
where $\kappa = k\sigma$ and the Stokes-Einstein relation
\addtocounter{eqnnum}{1}
\begin{equation}
D_0 = \frac{k_BT}{3 \pi \eta_0 \sigma}
\end{equation}
has been used.

Although the expression (25) for $\eta_N(\phi)$ has been derived for large 
$\phi \: ( 0.3 < \phi < 0.55)$, where cage diffusion is the dominant finite
time contribution to the viscosity (via eqs.(13) and (14)), eq.(25) nevertheless
appears to describe the $\phi$-dependence of $\eta_N(\phi)$ for small and
intermediate concentrations also, due to the presence of the $\eta_0 \chi(\phi)$
term (cf.fig.3). Fig.3 also shows that the cage diffusion describes the very rapid increase of 
$\eta_N(\phi)$ with $\phi$ for $0.40 < \phi < 0.55$ very well.

Eq.(25) has been evaluated using the Henderson-Grundke 
correction$^{[24]}$ to the Percus-Yevick equation for the computation of the
hard sphere $S_{eq}(k;\phi)$ and $S'_{eq}(k;\phi)$.  A convenient
Pad\'{e} approximation of $\eta_N(\phi)$ for practical use for all 
$0 < \phi < 0.55$ is: 
\addtocounter{eqnnum}{1}
\begin{equation}
\eta_N(\phi) = \eta_0 \chi(\phi) [1 + \frac{1.44 \phi^2 \chi(\phi)^2}
{1 - 0.1241 \phi + 10.46 \phi^2}]
\end{equation}
within a relative accuracy of less than 0.25\%. This approximation yields for
$\eta_N(\phi)$ the correct Einstein coefficient $\frac{5}{2} \phi$ as well as
the same coefficient of $O(\phi^2)$ as eq.(25).

Cichocki and Felderhof have obtained on the basis of the pair 
Smoluchowski
equation exact results for $\eta(\phi,\omega)$ to $O(\phi^2)$. Their result to
$O(\phi^2)$ for $\eta_N(\phi)$ is, without Brownian motion 
contributions$^{[25]}$:
\renewcommand{\theequation}{\arabic{eqnnum}\alph{eqnletter}}
\setcounter{eqnnum}{28}
\setcounter{eqnletter}{1}
\begin{equation}
\eta_N(\phi) = 1 + \frac{5}{2} \phi + 5.00 \phi^2
\end{equation}
while, with Brownian motion contributions they find$^{[26]}$:
\addtocounter{eqnletter}{1}
\begin{equation}
\eta_N(\phi) = 1 + \frac{5}{2} \phi + 5.91 \phi^2
\end{equation}
This can be compared with the approximate result we obtain from eq.(19):
\addtocounter{eqnletter}{1}
\begin{equation}
\eta_N(\phi) = 1 + \frac{5}{2} \phi + 6.03 \phi^2
\end{equation}
where the term $6.03 \phi^2$ contains a contribution $4.59 \phi^2$ from
$\eta_{\infty}(\phi)$ and a contribution $1.44 \phi^2$ from the second 
(mode-mode coupling) term between the square brackets on the r.h.s. of eq.(25).
Since for $\phi <0.25$ the cage-diffusion contribution to $\eta(\phi;\omega)$
can be neglected, eq.(22) then reduces to $\eta_N(\phi) = \eta_{\infty}(\phi) = 
\eta_0\chi(\phi)$.  The eqs.(28b) and (28c) both give then a very good 
representation of the experimental values for $\eta_N(\phi)$.

\section{Visco-elastic behavior}

For $\omega \neq 0$, $\eta(\phi, \omega)$ of eq.(20) is complex, so that the
visco-elastic behavior of the suspension can be written in the form:
\renewcommand{\theequation}{\arabic{eqnnum}}
\addtocounter{eqnnum}{1}
\begin{equation}
\eta(\phi, \omega) = \eta'(\phi, \omega) + i \eta''(\phi, \omega)
\end{equation}
where $\eta'(\phi, \omega)$, $\eta''(\phi, \omega)$ are the real and 
imaginary parts of $\eta(\phi, \omega)$, 
respectively.  It is convenient and customary$^{[5]}$ to consider, instead of
$\eta'(\phi, \omega)$ and $\eta''(\phi, \omega)$ reduced quantities defined by:
\renewcommand{\theequation}{\arabic{eqnnum}\alph{eqnletter}}
\setcounter{eqnnum}{30}
\setcounter{eqnletter}{1}
\begin{equation}
\eta^*_R(\phi,\omega) = \frac{\eta'(\phi,\omega)-\eta(\phi,\infty)}
{\eta(\phi,0)-\eta(\phi,\infty)} = \frac{\eta'(\phi,\omega)-\eta_\infty(\phi)}
{\eta_N(\phi)-\eta_{\infty}(\phi)} 
\end{equation}
and
\addtocounter{eqnletter}{1}
\begin{equation}
\eta^*_I(\phi,\omega) = \frac{\eta''(\phi,\omega)}
{\eta_N(\phi) - \eta_{\infty}(\phi)}
\end{equation}
where the reduced real part $\eta^*_R(\phi,\omega)$ varies as a function of
$\omega$ between 1 (for $\omega \rightarrow 0)$ and 0 (for $\omega \rightarrow
\infty)$ for all $\phi$ and $\eta^*_I(\phi, \omega)$ vanishes for 
$\omega \rightarrow 0$ and $\omega \rightarrow \infty$, exhibiting a maximum 
in between.  In fig.4, $\eta^*_R(\phi,\omega)$ and 
$\eta^*_I(\phi,\omega)$ are compared with the experimental data of Van der 
Werff et al, as a function of a reduced $\omega$ for all available 
$\phi$ for $0.44 \leq \phi \leq 0.57^{[5]}$.  As Van der Werff et al state, 
the values they find for the reduced quantities $\eta^*_R(\phi, \omega)$ and 
$\eta^*_I(\phi, \omega)$ are very weakly dependent on $\phi$, 
which is consistent with the crowding of all experimental points around
the theoretical curves, inside the experimental errors. 
The scaling of $\omega$ for the experimental data was performed in the 
same way as was done by Van der Werff et al by fitting the data for large
$\omega$ to the expression (cf.section 6):
\renewcommand{\theequation}{\arabic{eqnnum}}
\addtocounter{eqnnum}{1}
\begin{equation}
\eta^*_R(\phi,\omega) = \eta^*_I(\phi,\omega) = \frac{3\sqrt{2}}{2 \pi}
\frac{1}{\sqrt{\omega \tau_1(\phi)}}
\end{equation}
where $\tau_1(\phi)$ is a phenomenological time for the experiments.  The 
$\tau_1(\phi)$ used for the theoretical results is given in section 6, eq.(33).

Nevertheless a more detailed comparison of $\eta^*_{R,I}(\phi,\omega)$
as a function of $\phi$ can be made, although the large experimental
uncertainties of the data and the difference in the basic inputs in the
theory ($\phi$ and $\eta_0$) and experiment ($\sigma, c$ and $\eta_0$, with
$c$ the weight concentration of the colloidal particles)
complicate considerably a compelling detailed comparison of theory and 
experiment.  Examples are given in fig.5. In the same figure the results 
of a general phenomenological description of the visco-elastic behavior of 
colloidal suspensions due to Cichocki and Felderhof are given$^{[23]}$.  This 
description is based on a three pole approximation in the complex 
$\sqrt{\omega}$-plane, whose location is derived from the experimentally
measured values  $\eta_N^{exp}(\phi), \eta_{\infty}^{exp}(\phi)$ and three 
additional parameters, one of them being a relaxation time. From these three 
poles the $\eta'(\phi,\omega)$ and
$\eta''(\phi, \omega)$ as a function of $\omega$ can be derived.  For the
three concentrations $\phi = 0.44$, 0.46 and 0.53, for which their procedure 
could be implemented, $\eta'(\phi,\omega)$ and $\eta''(\phi,\omega)$ are 
consistent with our 
results within the experimental errors.  As was shown by Cichocki and
Felderhof, the strongly deviating cloud of points near $\omega \tau_1(\phi) 
\approx 1$ in the imaginary part of the reduced viscosity 
$\eta^*_I (\phi,\omega)$ (cf.fig.4b) can be disgarded, since they violate the 
Kramers-Kronig relations between the real and the imaginary part of 
$\eta(\phi,\omega)$ and must therefore be erroneous$^{[23]}$.\\
\section{Large $\omega$-behavior}  

For large $\omega$, eq.(20) for $\eta(\phi,\omega)$ can be written as:
\addtocounter{eqnnum}{1}
\begin{equation}
\eta(\phi,\omega) = \eta_{\infty}(\phi) + \frac{9}{5} \phi^2 \chi^{5/2}
\eta_0 \frac{1}{\sqrt{\omega \tau_P}} (1 + i) + O (\frac{1}{\omega})
\end{equation}
where the square root singularity for $\omega \rightarrow \infty$ is induced
by the large $k$-behavior of the integrand on the r.h.s. of eq.(20), as given 
by eq.(21).  We note that the correction $O(\frac{1}{\omega})$ is an exact
result for low concentrations to $O(\phi^2)$ (cf. Appendix A) and is consistent
with what is found in the mode-mode coupling approximation.

Using eq.(32) in eq.(30) and comparing with eq.(31) gives for $\tau_1(\phi)$
the theoretical expression:
\addtocounter{eqnnum}{1}
\begin{equation}
\tau_1(\phi) = \frac{25}{18\pi^2\phi^4 \chi(\phi)^5} 
[\frac{\eta_N(\phi)}{\eta_0} - \chi(\phi)]^2 \tau_P
\end{equation}
which is plotted in fig.6 and is consistent with the experimentally used 
$\tau_1(\phi)$ up to about $\phi \approx 0.55$, averaging at a value of 
about $\tau_P/4$ (cf. section IV.B in ref.5).
The systematically too low theoretical value of 
$\tau_1(\phi)$ corresponds to the systematically too high theoretical value 
of the coefficient of the $\omega^{-1/2}$-singularity in eq.(32) as compared
with the exact value given in eq.(41) below.

In fact, in order to investigate this behavior further, an independent 
evaluation of
$\eta(\phi,\omega)$ for large $\omega$ was made, starting from a Green-Kubo
like formula for $\eta(\phi,\omega)$ rather than from eq.(8): 
\addtocounter{eqnnum}{1}
\begin{equation}
\eta(\phi,\omega) = \eta_{\infty}(\phi) + \frac{\beta}{V} \int^{\infty}_0 dt
\rho_\eta (t;\phi) e^{i \omega t}
\end{equation}
Here the stress-stress auto correlation function $\rho_{\eta}(t)$ is defined 
by:
\addtocounter{eqnnum}{1}
\begin{equation}
\rho_{\eta}(t;\phi)=< \Sigma^{\eta}_{xy} e^{\Omega t} \Sigma^{\eta}_{xy} >_{eq}
\end{equation}
where the brackets denote an equilibrium ensemble average.  Here, instead of
using the microscopic pressure tensor (the expression within the square
brackets of eq.(3) in section 2), we use the in this context more customary
microscopic stress tensor $\Sigma_{xy}^{\eta}$, which is equal but opposite in
sign and can be written as:
\addtocounter{eqnnum}{1}
\begin{equation}
\Sigma^{\eta}_{xy} = \sum^N_{i=1} r_{i,x} {\em{F}}_{i,y}
\end{equation}
with ${\bf{F}}_i = - \nabla_i \Phi (r^N)$ the force on particle $i \: (\nabla_i 
= \partial/\partial {\bf{r}}_i), \Phi(r^N) = \sum^N_{i<j=1} V(r_{ij})$ the 
total potential energy of the colloidal particles and 
\addtocounter{eqnnum}{1}
\begin{equation} 
\Omega = D_s \sum^N_{i=1} [\nabla_i + \beta F_i] \cdot \nabla_i
\end{equation}
the $N$-particle Smoluchowski operator$^{[27]}$ with $D_0$ replaced by the
short time self-diffusion coefficient $D_s(\phi)$
to make eq.(34) applicable to all fluid densities. This is further
discussed below.  For $N = 2$
and $\chi(\phi) = 1$ the adjoint operator occurs in the pair Smoluchowski equation 
(eq.(9)).  

The short time behavior of $\rho_\eta (t;\phi)$ determines the 
large $\omega$ behavior of $\eta(\phi,\omega)$.  Since for hard spheres
the interparticle potential is singular, one determines the short time
behavior of $\rho_\eta(t;\phi)$ by first using a soft potential  
$V_{\ell}(r) = \epsilon (\frac{\sigma}{r})^{\ell}$, where $\epsilon$ is 
the two particle interaction energy for $r = \sigma$, and then letting
$\ell \rightarrow \infty$, so that $V_{\ell}(r)$ approaches a potential
between two hard spheres of diameter $\sigma$.  For $\ell \rightarrow \infty$,
one can then derive for $\rho_\eta(t,\phi)$ the expression$^{[28]}$:
\renewcommand{\theequation}{\arabic{eqnnum}\alph{eqnletter}}
\setcounter{eqnnum}{38}
\setcounter{eqnletter}{1}
\begin{equation}
\rho_\eta(t;\phi) = \frac{2\pi n^2 V \sigma^3 \chi(\phi) \ell}{15 \beta^2} 
r(t^*)
\end{equation}
with
\addtocounter{eqnletter}{1}
\begin{equation}
r(t^*) = \int^{\infty}_0 dse^{-s} e^{t^*[(s^2 \frac{\partial}{\partial s} +
s - s^2) \frac{\partial}{\partial s}]} s
\end{equation}
where
\addtocounter{eqnletter}{1}
\begin{equation}
t^* = \frac{2 D_s t \ell^2}{\sigma^2}
\end{equation}
The leading term of $r(t^*)$ for $\lim_{t \rightarrow 0} \lim_{l \rightarrow
\infty}$, i.e., $t^* \sim tl^2 \rightarrow \infty$, which determines 
the short time behavior of $\rho_{\eta}(t;\phi)$ for a hard-sphere potential,
was obtained by M. J. Feigenbaum and reads$^{[28]}$: 
\renewcommand{\theequation}{\arabic{eqnnum}}
\addtocounter{eqnnum}{1}
\begin{equation}
r(t^*) = \frac{1}{\sqrt{\pi t^*}}
\end{equation}
Using eqs.(34), (38) and (39) and the Stokes-Einstein relation (26), one
obtains for $\eta(\phi,\omega)$ for large
$\omega$ and for a hard sphere potential for all $\phi$ the exact expression:
\addtocounter{eqnnum}{1}
\begin{equation}
\eta(\phi,\omega) \sim \eta_{\infty}(\phi) + \frac{18}{5} \phi^2\chi(\phi)
\eta_0 \left[ \frac{D_0}{D_s(\phi)} \right]^{1/2}
\frac{1+i}{\sqrt{\omega \tau_P}}
\end{equation}
Using then that $D_s(\phi) = D_0/\chi(\phi)$ (cf. section 8, sub. 3) one has:
\addtocounter{eqnnum}{1}
\begin{equation}
\eta(\phi,\omega) \sim \eta_{\infty}(\phi) + \frac{18}{5} \phi^2\chi(\phi)^{3/2}
\eta_0 \frac{1}{\sqrt{\omega \tau_P}} (1 + i)
\end{equation}
Eqs.(32) and (41) are both compared with the 
experimental data for large $\omega$ and for most experimental values of 
$\phi$ in fig.7. We emphasize that in order to get agreement with experiment it
is necessary to replace the low density Stokes-Einstein diffusion coefficient
$D_0$ by the self-diffusion coefficent $D_s(\phi)$ in the
basic Smoluchowski operator (cf.eq.(37) and fig.7). We also emphasize that
the exact result of eq.(41) constitutes a generalization of Cichocki and 
Felderhof's low concentration result to all concentrations in the fluid
range.  
A detailed derivation of eq.(41) will be given elsewhere$^{[28]}$.

It is clear that the experiments agree very well with eq.(41)
and not with eq.(32), consistent with the systematically lower theoretical
values of $\tau_1(\phi)$ in fig.6. This could well be related to the 
approximations made to obtain eq.(32): (1) the use of the complete 
$S_{eq}(k;\phi)$ (i.e. for all $\phi$) in the two particle
eq.(15) and the  use of $\omega_H(k;\phi)$ as the only basic relaxation time;
(2) the replacement of the potential $V(r)$ in eq.(8) by the direct 
correlation function $C_{eq}(r;\phi)$ and (3) the neglect of
the force term on the l.h.s. of eq.(10) and consequenly 
the correct boundary condition of hard sphere impenetrability incurred by
the Fourier transform from eq.(10) to eq.(12) (cf.Appendix A).

The first approximation was intended to incorporate
in the calculation of $\eta(\phi,\omega)$ contributions due to more than two 
isolated particles, i.e., correcting for the neglect of the three particle
distribution function in the eq.(9) for $g(r;\phi,\omega,\gamma_0,t)$.

As pointed out before, the second approximation, is necessary to perform a
Fourier transform of eq.(8). It also corrects partly for the unphysical 
contributions from overlapping particle configurations, due to the neglect
of the proper hard sphere boundary condition (cf.Appendix A). We remark that 
the Fourier transform of eq.(8) was due to the necessity of 
introducing the relaxation times $\tau_c(k;\phi)$ related to the cage 
diffusion for concentrated colloidal suspensions, which have only been 
determined for periodic particle arrangements, characterized by a wave number 
$k$. However, neither of these two approximations seem to be responsible for 
the incorrect asymptotic $\omega$-behavior of $\eta(\phi,\omega)$.

As to the third approximation,
if we compare eqs.(32) for low densities, i.e. $\chi(\phi)
= 1$, with the exact solution for $\eta(\phi,\omega)$ obtained by 
Cichocki and Felderhof$^{[16]}$ to $O(\phi^2)$, we see that the second term on
the r.h.s. of eq.(32) is smaller by a factor 2.
Cichocki and Felderhof considered eq.(10) with the correct hard sphere boundary 
condition in ${\bf{r}}$-space and solved it exactly for all $t$.  If we solve
eq.(10) in the same manner but neglect the force term on the l.h.s. 
(cf.Appendix A), we obtain, however, eq.(32) in the limit of large $\omega$ 
with $\chi(\phi) =1$.
This suggests that the third approximation, the neglect of the force term on 
the l.h.s. of eq.(10) and the ensuing violation of the proper hard sphere 
boundary condition in real space in making the Fourier transform from eq.(10) 
to eq.(12) is the main reason for the erroneous expression (32).

We note that the eqs.(32) and (41) show that the difference between the exact 
and the mode coupling result for the coefficient of $\omega^{-1/2}$ 
is a constant factor $2/\chi(\phi)$. This only affects the approach to 
$\omega = \infty$, not $\eta_{\infty}(\phi)$ itself, and is of no influence 
if one plots the mode coupling theory
on the phenomenological time-scale $\omega \tau_1(\phi)$ using eq.(33) 
(cf.fig.5).  This may be of practical importance for predicting the 
visco-elastic behavior of concentrated colloidal suspensions since the scaling
in time does not affect the Newtonian behavior of the viscosity
$^{[29]}$.
 
\section{Small $\omega$-behavior}

For low densities to $O(\phi^2)$ the small $\omega$, or long time, behavior 
of $\eta(\phi,\omega)$ follows from eqs.(20), (21) and (29) to be:
\renewcommand{\theequation}{\arabic{eqnnum}\alph{eqnletter}}
\setcounter{eqnnum}{42}
\setcounter{eqnletter}{1}
\begin{equation}
\frac{\eta'(\phi,\omega) - \eta_{\infty}(\phi)}{\eta_0}= \{\frac{36}{25} - 
\frac{32}{175} (\omega \tau_{\em{P}})^2\}\phi^2 + ....
\end{equation}
\addtocounter{eqnletter}{1}
\begin{equation}
\frac{\eta''(\phi,\omega)}{\eta_0} = \frac{48}{175}(\omega \tau_{\em{P}})
\phi^2 + ....
\end{equation}
This can be compared with the exact results of Cichocki and
Felderhof$^{[16]}$ to $O(\phi^2)$ for $\omega \rightarrow 0$:
\addtocounter{eqnnum}{1}
\setcounter{eqnletter}{1}
\begin{equation}
\frac{\eta'(\phi,\omega)-\eta_{\infty}(\phi)}{\eta_0} = \{ \frac{12}{5} - 
\frac{16}{81} (\omega \tau_{\em{P}})^2 \} \phi^2 + ....
\end{equation}
\addtocounter{eqnletter}{1}
\begin{equation}
\frac{\eta''(\phi,\omega)}{\eta_0} = \frac{8}{15} \phi^2 (\omega \tau_{\em{P}})
+ ....
\end{equation}
The agreement of eqs.(42a,b) with eqs.(43a,b) for small $\omega$ and low 
concentrations, in particular of the coefficient of $(\omega \tau_P)^2$ in 
the real parts, is better than that of eq.(32) and eq.(41) for 
large $\omega$. This is probably due to the fact that the neglect of the 
proper hard sphere boundary condition in the mode-mode coupling theory is more
serious for a description of the short time behavior than the long time 
behavior of the 
suspension. We remark however that the difference in the first terms on the
r.h.s. of the eqs.(42a) and (43a), i.e. 36/25 and 12/5, respectively, is a
direct consequence of the violation of the proper hard sphere boundary 
condition (cf.Appendix A, in particular eq.(A.25))
\section{Discussion}

1. The $\omega$-dependence of $\eta(\phi,\omega)$ is well represented by 
eq.(20) for all $\phi$ on the phenomenological time-scale $\tau_1(\phi)$ or
if plotted as a function of $\omega \tau_P$, when an over-all shift to the 
theoretical curves of $2/ \chi(\phi)$ is applied$^{[29]}$. This is due to the 
fact that the asymptotic mode-mode coupling result (32) for the large
$\omega$ behavior of $\eta(\phi,\omega)$ is not correct, because of the 
incomplete incorporation of the hard sphere impenetrability in the theory.
The mode-mode coupling contribution to $\eta(\phi,\omega)$ 
should be best for values of $\omega$ around $\omega \tau_1(\phi) \approx 1$, 
where there are rather few experimental points.
It would be interesting therefore if a more detailed comparison between theory
and experiment could be made in this $\omega$-regime, to obtain a more 
appropriate test for the validity of the mode-mode coupling theory used here.

2. The result (20) for $\eta(\phi,\omega)$ is based exclusively on the
instantaneous time behavior of $\eta_{\infty}(\phi)$ and the cage-diffusion
relaxation mechanism.  From the agreement of $\eta(\phi,\omega)$ and
$\eta_N(\phi)$ with experiment, 
it would seem that these two physical processes essentially suffice to 
understand the Newtonian as well as the viscous behavior in the entire fluid 
range of hard sphere colloidal suspensions.  That this agreement occurs without
considering explicitly any hydrodynamical interactions between the colloidal
particles
in the theory presented here may appear rather puzzling.  We do not have an explanation
for this, other than that at high concentrations, where $0.3 < \phi < 0.55$, 
the surface to surface distance between the hard spheres is so small, that
a ``quenching'' of hydrodynamical effects is not unthinkable.

3. There may, however, be a deeper justification for the neglect of the usual
hydrodynamical interactions in our theory.  It seems that in a number of 
cases the same dependence of a physical quantity of the suspension can be
obtained by theories with and without hydrodynamical interactions between
the Brownian particles. In this respect the following two observations are 
relevant.

(a) The concentration dependence of the infinite frequency viscosity
$\eta_{\infty}(\phi)$ as well as of the Newtonian viscosity $\eta_N(\phi)$ for
low and intermediate concentrations $0 \leq \phi \leq 0.45$ are described by 
our relations 
(cf.eqs.(22) and (25)):
\addtocounter{eqnnum}{1}
\setcounter{eqnletter}{1}
\begin{eqnarray}
\eta_{\infty}(\phi) & = & \eta_0 \chi(\phi) = \nonumber \\
& = & \eta_0 [ 1 +\frac{5}{2} \phi + 4.59 \phi^2 + O(\phi^3)]
\end{eqnarray}
and
\addtocounter{eqnletter}{1}
\begin{eqnarray}
\eta_N(\phi) & = & \eta_0 \chi (\phi) [1 + \frac{1}{40\pi} \int^{\infty}_0
d \kappa \kappa^2 \frac{[S'_{eq}(\kappa;\phi)]^2}{S_{eq}(\kappa;\phi)
d(\kappa)}] = \nonumber \\
& = & \eta_0 [ 1 + \frac{5}{2} \phi + 6.03 \phi^2 + O(\phi^3)]
\end{eqnarray}
respectively. The r.h.s. of eqs.(44a) and (44b) can be compared with 
Beenakker's expression$^{[30]}$:
\addtocounter{eqnletter}{1}
\begin{eqnarray}
\eta^{eff}(\phi) & = & {\rm{lim}}_{k \rightarrow 0} [\eta(k;\phi)] = \nonumber
\\ & = & \eta_0 [ 1 + \frac{5}{2} \phi + 4.84 \phi^2 + O(\phi^3)]
\end{eqnarray}
for, what he calls, the effective viscosity. Beenakker's $\eta^{eff}(\phi)$ 
is derived from a wave vector dependent viscosity $\eta(k;\phi)$, a complicated 
functions of $k$, by using the quasi-static 
Stokes equation to describe the motion of the fluid, 
neglecting inertial effects. This implies, as he pointed out, that his 
equation is valid 
for $\tau_B < t < \tau_P$. Our relations (44a) and (44b), however, are valid 
for $t < \tau_B$ and $t > \tau_P$, respectively. Thus his result (eq.(44c)) 
can be regarded as between eq.(44a) and eq.(44b) (cf.fig 8a). While for low 
concentrations the difference between the three expressions (as well as 
eqs.(28a) and (28b)) is marginal, since it does not
appear to be relevant for comparison with experiment, we emphasize
that the strong experimental increase of the Newtonian viscosity 
for higher concentrations $\phi > 0.3$, can only be described by the
integral on the r.h.s. of eq.(44b) (cf.figs.2 and 8a).

(b) Also, the concentration dependence of the short time self-diffusion coefficient 
$D_s(\phi)$ for low and intermediate concentrations
$0 \leq \phi \leq 0.45$ can be equally well described, within the 
experimental uncertainties, by our relation:
\addtocounter{eqnnum}{1}
\setcounter{eqnletter}{1}
\begin{equation}
D_s(\phi) = \frac{D_0}{\chi(\phi)}
\end{equation}
as by the Beenakker and Mazur expression$^{[31]}$:
\addtocounter{eqnletter}{1}
\begin{equation}
D_s(\phi) =  {\rm{lim}}_{k \rightarrow \infty} D(k;\phi)
\end{equation}
where $D(k;\phi)$ is a wave vector dependent collective diffusion coefficient,
which is, like $\eta(k;\phi)$, a complicated function of $k$.
While our relation (45a) for $D_s(\phi)$ is valid for 
$t < \tau_B$, Beenakker and Mazur's expression (45b) is, like for their 
viscosity, valid for $\tau_B < t < \tau_P$.  On this larger time-scale 
$D_s(\phi)$ will contain extra, in his case, hydrodynamic contributions in 
addition to our instantaneous contributions, leading to slightly 
larger values for the short time self-diffusion coefficient. The same
obtains for the experiments of van Megen et al$^{[45]}$ and
Pusey and van Megen$^{[46]}$ (cf.fig 8b).

Beenakker and Mazur consider only purely hydrodynamic interactions between 
the particles, in that they study the hydrodynamical effect of a number of 
stationary particles on the motion of one moving particle. In our 
case no hydrodynamics enters explicitly at all, essentially only molecular 
considerations are used. For short times the
(static) equilibrium radial distribution at contact $\chi(\phi)$, derived from
the canonical distribution of the colloidal particles in equilibrium, occurs,
yet a comparable agreement with experiment is obtained. 
For long times there is an extra (dynamic) contribution due to the 
increasing difficulty for a particle to diffuse out of the cage formed by its
neighbors.

(c) We believe that for a complex system like a colloidal suspension
there could be apparently very different alternate descriptions
of the same phenomena.  Perhaps the simplest and most striking example of 
this is the observation that Einstein's low concentration result for the
viscosity of a colloidal suspension, derived from Stokes 
hydrodynamics$^{[32]}$ 
\addtocounter{eqnnum}{1}
\setcounter{eqnletter}{1}
\begin{equation}
\frac{\eta_{\infty}(\phi)}{\eta_0} = 1 + \frac{5}{2} \phi + O(\phi^2)
\end{equation}
can also be obtained, using an Einstein relation (cf.eqs.(44a) and (45a)):
\addtocounter{eqnletter}{1}
\begin{equation}
\frac{\eta_{\infty}(\phi)}{\eta_0} = \frac{D_0}{D_s(\phi)} = 1 + \frac{5}{2} 
\phi + O(\phi^2)
\end{equation}
Although these equivalent alternate descriptions of colloidal suspension
properties - and especially eq.(46b) - could well be a fluke, a deeper
origin cannot be ruled out in our opinion either.

In fact, for the equivalence of Einstein's expression (46a) and our (46b)
the following physical argument can be given.

Felderhof has shown$^{[33]}$ - and it also follows from the 
Green-Kubo expression (34) - that $\eta(\phi,\omega) = \eta_0 [1 + \frac{5}{2} 
\phi + \eta_2(\omega) \phi^2]$.  Therefore the first two terms in the 
expansion of $\eta (\phi,\omega)$ in powers of $\phi$ are independent of 
$\omega$.  This implies that when computed for any $\omega$ they should 
give the same answer: $\eta_0 [1 + \frac{5}{2} \phi]$.

Einstein - as represented in Landau-Lifshitz$^{[34]}$ - did the computation
for $\omega = 0$, i.e., he used a long time stationary state hydrodynamic
calculation, to obtain the extra resistance of the suspension to shear from 
the change of the velocity field of the fluid due to a single Stokesian hard
sphere particle placed in it.

We propose to do a computation at $\omega = \infty$, i.e., for a very short
(in fact, instantaneous) time.  Then the placing of one particle - or even many
mutually separated particles - in the solvent will not have any effect on the
viscous resistance of the suspension.  The only way the presence of the 
particles can produce an extra flow resistance is from pairs of particles 
(already) in 
contact, where an ``instantaneously'' collision takes place adding to the
viscous dissipation in the suspension.  Therefore, for $\omega = \infty$ 
the increase
in the effective fluid viscosity as a function of $\phi$ will be given by the 
relative increase in the number of particle pairs at contact in equilibrium 
as a function of $\phi$, which is $\chi(\phi)$.  On the basis of this argument
one would conjecture that for $\omega = \infty$, the increase in suspension
viscosity, when compared with that of the pure solvent, would be $\chi(\phi)$
for all $\phi$, not just $1 + \frac{5}{2} \phi$ to $O(\phi)$.  This 
conjecture is consistent with experiment (as shown in fig.2) and should
be derivable from kinetic theory$^{[35]}$.

4. We also remark that the Einstein relation
\addtocounter{eqnnum}{1}
\setcounter{eqnletter}{1}
\begin{equation}
D_0 = \frac{k_BT}{3 \pi \eta_0 \sigma}
\end{equation}
appears to hold not only for infinitely dilute suspensions, but for all 
concentrations in the form$^{[4]}$:
\addtocounter{eqnletter}{1}
\begin{equation}
D_s(\phi) = \frac{D_0}{\chi(\phi)}=\frac{k_BT}{3 \pi \eta_{\infty}(\phi)\sigma}
\end{equation}
as can be seen in fig.8c. The physical reason for this seems to be that as 
long as the times of
observation are sufficiently short (or the frequencies sufficiently
high), so that no significant motion of the colloidal particles can take place,
no hydrodynamical effects will occur, and only the instantaneous effect due
to particles at contact - which does not require any time to occur -, i.e.,
$\chi(\phi)$ will be relevant.

5. Recently Brady$^{[36]}$ has published a different model for the Newtonian
as well as the frequency dependent viscosity. His results can be
obtained from the low density result of Cichocki and Felderhof$^{[16]}$ 
(cf.Appendix A) with only two modifications: (1) a scaling of their exact
solution (eqs.(A.2) and (A.6)) for the low density two particle Smoluchowski
eq. (10) (eq.(A.1)), by replacing the Stokes-Einstein diffusion
coefficient $D_0$ by the short time selfdiffusion coefficient $D_s(\phi)$
and (2) the addition of a factor $g_{eq}(r=\sigma;\phi) = \chi(\phi)$
to the low density expression for the potential contribution of the 
viscosity in terms of the pair distribution function (cf. the second term on
the r.h.s. of eq.(8)). This leads directly to Brady's expression for
$\eta(\phi,\omega)$ (cf.eq.(A.11), which in our notation reads:
\renewcommand{\theequation}{\arabic{eqnnum}}
\setcounter{eqnnum}{48}
\begin{equation}
\eta(\phi,\omega) = \eta_{\infty}(\phi) + \eta_0 \phi^2 \alpha_V (\omega)
g_{eq}(\sigma;\phi) \frac{D_0}{D_s(\phi)}
\end{equation}
which reduces for $\omega = 0$ (with eq.(A.12)) to his expression for the
Newtonian viscosity $\eta_N(\phi)$:
\addtocounter{eqnnum}{1}
\begin{equation}
\eta_N(\phi) = \eta_{\infty}(\phi) + \frac{12}{5} \eta_0 \phi^2 g_{eq}
(\sigma;\phi) \frac{D_0}{D_s(\phi)}
\end{equation}
However, in his calculations Brady determines the
three basic ingredients of his theory empirically: $\eta_{\infty}(\phi)$ is 
derived from measurements and Stokesian dynamics$^{[37,38]}$, while 
$g_{eq}(\sigma;\phi)$ is taken to be given
by the Carnahan-Starling approximation eq.(24) for $0 < \phi < 0.5$ and by 
$1.2(1-\phi/ \phi_m)^{-1}$ for $\phi> 0.5$, as derived from dense h.s.
fluid computer simulations, where 
$\phi_m = 0.63$ is the volume fraction
of random close packing of hard spheres. Furthermore the relative short time 
self diffusion coefficient $D_s(\phi)/D_0$ is taken from Ladd's computer 
simulations for $0 < \phi < 0.45^{[38]}$ and from Phung's Stokesian dynamics 
simulations for $\phi > 0.45^{[39]}$. This leads to a curve for $\eta_N(\phi)$, 
as given by eq.(49), which is virtually indistinguishable from our $\eta_N(\phi)$
based on eq.(25) for $0 < \phi < 0.55$. We remark that eq.(49), with the
just mentioned determination of $\eta_{\infty}(\phi)$, $g_{eq}(\sigma;\phi)$
and $D_s(\phi)/D_0$, also describes
very well the experimental data for $\eta_N(\phi)^{[5,6,40,41]}$ for
$0.55 < \phi < 0.60$, where the precise thermodynamic state of the suspension
is not clear, while eq.(25) gives then too low values for $\eta_N(\phi)$. 
Virtually the same result as Brady's
description of $\eta_N(\phi)$ for $0 < \phi < 0.60$ can be obtained by using
in his eq.(49) for all $\phi$, our eqs.(22) and (45a) for 
$\eta_{\infty}(\phi)$ and $D_s(\phi)/D_0$, respectively, as well as his 
representation of $g_{eq}(\sigma;\phi)$. It is clear that the precipitous 
increase of 
$\eta_N(\phi)$ for $\phi>0.55$ is then a direct consequence of the pole in 
$g_{eq}(\sigma;\phi)$ at $\phi = \phi_m$.

However, for the visco-elastic behavior, when plotted as a function of
$\omega \tau_1(\phi)$, Brady's results do not agree as
well with the experiments of Van der Werff et al$^{[36,42]}$. This may well be 
related to the fact that the basic ingredient of Brady's theory that causes the
increase of $\eta_N(\phi)$ for large $\phi$ is a static one, related to the 
behavior of $g_{eq}(\sigma;\phi) \sim (1 - \phi / \phi_m)^{-1}$ as 
random close packing is approached,
while in our theory it is a dynamic one: the increasing difficulty of diffusion
of a particle out of the cage formed by its neighbors. It appears 
that only the latter one is able to account for the frequency behavior of
$\eta(\phi;\omega)$. The underlying physics of the two processes is therefore 
very different: while we use the typical high density mechanism of cage diffusion,
Brady upgrades the low density physics by effectively  scaling with 
$g_{eq}(\sigma;\phi)$ and $D_s(\phi)$.

We note that essentially the same mode-mode coupling term as in eq.(25) gives 
the steep
viscosity rise at high densities for atomic liquids, since the atoms -
like the colloidal particles - find themselves in cages, out of which they can
only escape with increasing difficulty with increasing density$^{[19,20]}$.

{\Large{\bf{Acknowledgement}}}  E. G. D. C. gratefully acknowledges 
support from the U. S. Department of Energy under contract number DE-FG02-88-
ER13847 and R. V. support from the Netherlands Foundation for Fundamental
Research of Matter (FOM).
\newpage
\noindent {\underline{Appendix A}}

Here we compare for low densities $\phi \rightarrow 0$ and hard spheres the
exact dynamic viscosity $\eta(\phi,\omega)$ as obtained from eqs.(8) and (10)
by Cichocki and Felderhof$^{[16]}$ with the mode-mode coupling approximation
$\eta_{mc}(\phi,\omega)$ given by eq.(20).

We first give the exact solution 
of eq.(10) for $\delta g({\bf{r}};\phi,\omega)$ as obtained by Cichocki and
Felderhof.  For $\phi \rightarrow 0$, $g_{eq}(r; \phi) = 
{\rm{exp}} (- \beta V(r))$, so that eq.(10) reads:
\newcounter{chanum}
\newcounter{eqnnum1}
\renewcommand{\theequation}{A.\arabic{eqnnum1}}
\setcounter{eqnnum1}{1}
\begin{equation}
[ -i \omega + 2 D_0 {\bf{\nabla}} \cdot \{\beta {\bf{F}}({\bf{r}}) - 
{\bf{\nabla}}\}] \delta g ({\bf{r}};\phi,\omega) = \beta \frac{xy}{r} V'(r) 
e^{- \beta V(r)}
\end{equation}
with $V'(r) = \partial V(r)/\partial r$.  The solution of eq.(A.1) can be 
written as
\addtocounter{eqnnum1}{1}
\begin{equation}
\delta g({\bf{r}};\phi,\omega) = \frac{xy}{r^2} f (\frac{r}{\sigma};\omega)
e^{- \beta V(r)}
\end{equation}
Substitution of (A.2) into (A.1) and using that 
\addtocounter{eqnnum1}{1}
\begin{equation}
\{\beta {\bf{F}}({\bf{r}}) - {\bf{\nabla}}\} e^{- \beta V(r)} = 0
\end{equation}
one obtains in the hard sphere limit $V(r) = \lim_{l \rightarrow \infty} V_l(r)
= \lim_{l \rightarrow \infty} \epsilon (r/\sigma)^l$ the following 
equation for $f(u;\omega)$, with $u = r/\sigma$, 
\addtocounter{eqnnum1}{1}
\begin{equation}
[\frac{\partial}{\partial u} u^2 \frac{\partial}{\partial u} - 6 +
\frac{i\omega \sigma^2}{2 D_0} u^2] f(u;\omega) = 0
\end{equation}
with the boundary condition
\addtocounter{eqnnum1}{1}
\begin{equation}
f' (1;\omega) = \frac{\sigma^2}{2 D_0}
\end{equation}
where $f' (u;\omega) = \partial f(u;\omega)/\partial u$.  This boundary
condition ensures that the r.h.s. of (A.1), which diverges at
$r = \sigma$ for hard spheres, cancels exactly a similar divergent term 
arising from ${\bf{F}}({\bf{r}})$ on the l.h.s.  The solution of (A.4) with 
(A.5) is, for $r \geq \sigma (u \geq 1)$.
\addtocounter{eqnnum1}{1}
\begin{equation}
f (u;\omega) = \frac{\sigma^2}{2 D_0} \: \frac{k_2(\alpha u)}
{\alpha k'_2 (\alpha)}
\end{equation}
with $k_2(x)$ the modified spherical Bessel function$^{[17]}$ of the third kind,
\addtocounter{eqnnum1}{1}
\begin{equation}
k_2(x) = e^{-x} \{x^{-1} + 3 x^{-2} + 3 x^{-3}\}
\end{equation}
and 
\addtocounter{eqnnum1}{1}
\begin{equation}
\alpha = \alpha(\omega) = (1 - i) \sqrt{\frac{\omega \sigma^2}{4 D_0}}
\end{equation}
We note that for hard spheres $f (r/\sigma; \omega)$ is continuous at 
$r = \sigma$ so that $\delta g ({\bf{r}};\phi,\omega)$ in (A.2) shows a jump 
at $r = \sigma$ due to the factor ${\rm{exp}}(- \beta V(r)) = 
\Theta (r - \sigma)$ with
$\Theta (x)$ the Heaviside step function.  In particular, $\delta g ({\bf{r}};
\phi,\omega) = 0$ for $r < \sigma$, reflecting the inpenetrability 
of two hard spheres.  Next we substitute (A.2) for $\delta g ({\bf{r}}; \phi,
\omega)$ in eq.(8) for $\eta(\phi;\omega)$.  Using that for hard spheres
\addtocounter{eqnnum1}{1}
\begin{equation}
V'(r) e^{-\beta V(r)} = -k_BT \delta(r - \sigma)
\end{equation}
one obtains straightforwardly
\addtocounter{eqnnum1}{1}
\begin{equation}
\eta(\phi;\omega) = \eta_{\infty}(\phi) - \frac{2 \pi}{15} k_BT n^2 \sigma^3
f(1;\omega)
\end{equation}
Substitution of (A.6) and (A.7) leads to the final result for 
$\phi \rightarrow 0$,
\addtocounter{eqnnum1}{1}
\begin{equation}
\eta(\phi;\omega) = \eta_{\infty}(\phi) + \eta_0 \phi^2 \alpha_V(\omega)
\end{equation}
with
\addtocounter{eqnnum1}{1}
\begin{equation}
\alpha_V(\omega) = \frac{36}{5} \frac{\alpha^2 + 3 \alpha + 3}
{\alpha^3 + 4 \alpha^2 + 9 \alpha + 9}
\end{equation}
and $\alpha = \alpha(\omega)$ given by (A.8).

In the mode-mode coupling theory on the other hand, one neglects the
force ${\bf{F}}({\bf{r}})$ on the l.h.s. of (A.1), so that
$\delta g_{mc}({\bf{r}}; \phi,\omega)$ satifies  
\addtocounter{eqnnum1}{1}
\begin{equation}
[-i\omega -2 D_0 \nabla^2] \delta g_{mc}({\bf{r}};\phi,\omega) = \beta
\frac{xy}{r} V'(r) e^{-\beta V(r)}
\end{equation}
The solution of this equation can be written in the form:
\addtocounter{eqnnum1}{1}
\begin{equation}
\delta g_{mc}({\bf{r}};\phi,\omega) = \frac{xy}{r^2} f_{mc} 
(\frac{r}{\sigma};\omega)
\end{equation}
Substitution of (A.14) into (A.13) yields the following equation for
$f_{mc}(u;\omega)$:
\addtocounter{eqnnum1}{1}
\begin{equation}
[\frac{\partial}{\partial u} u^2 \frac{\partial}{\partial u} - 6 + 
\frac{i \omega \sigma^2}{2 D_0} u^2] f_{mc} (u;\omega) = 0
\end{equation}
with boundary condition $(\epsilon \rightarrow 0)$
\addtocounter{eqnnum1}{1}
\begin{equation}
f'_{mc}(1 + \epsilon;\omega) - f'_{mc}(1 - \epsilon;\omega) = 
\frac{\sigma^2}{2 D_0}
\end{equation}
which follows from the r.h.s. of (A.13) in the hard sphere limit, using (A.9).
Thus, $f_{mc}(r/\sigma;\omega)$ is continuous for all $r$ with a jump in
its derivative at $r =  \sigma$ given by (A.16).  The solution of (A.15) and
(A.16) is for $u \leq 1$,
\addtocounter{eqnnum1}{1}
\begin{equation}
f_{mc}(u;\omega) = \frac{\sigma^2}{2 D_0} \: \frac{-K(\alpha)}{1 + K(\alpha)}
\: \frac{i_2(\alpha u)}{\alpha i'_2(\alpha)}
\end{equation}
and for $u \geq 1$,
\addtocounter{eqnnum1}{1}
\begin{equation}
f_{mc}(u;\omega) = \frac{\sigma^2}{2 D_0} \: \frac{1}{1 + K(\alpha)}
\: \frac{k_2(\alpha u)}{\alpha k'_2(\alpha)}
\end{equation}
where $\alpha = \alpha(\omega)$ is defined in (A.8), $k_2(x)$ in (A.7),
$i_2(x)$ is the modified spherical Bessel function of the second kind$^{[17]}$,
\addtocounter{eqnnum1}{1}
\begin{equation}
i_2(x)=(\frac{3}{x^3}+\frac{1}{x}){\rm{sin h}} x-\frac{3}{x^2} {\rm{cos h}} x
\end{equation}
and 
\addtocounter{eqnnum1}{1}
\begin{equation}
K(\alpha) = - \frac{k_2(\alpha) i'_2(\alpha)}{k'_2(\alpha) i_2(\alpha)} 
\end{equation}
Thus $\delta g_{mc}({\bf{r}};\phi,\omega)$ given by (A.14), (A.17) and (A.18)
is continuous for all $r$ and nonvanishing for $r < \sigma$, allowing two 
spheres to overlap.  To exclude such unphysical configurations in eq.(8) for
the viscosity $\eta(\phi,\omega)$ we replace $V(r)$ by $-k_BT C_{eq}(r;\phi)$
(eq.(17)).  Using that $C_{eq}(r;\phi) = {\rm{exp}}(-\beta V(r)) - 1$ for
$\phi \rightarrow 0$, $\partial V(r)/\partial x$ in eq.(8) is then replaced
by
\addtocounter{eqnnum1}{1}
\begin{equation}
\frac{\partial V(r)}{\partial x} \longrightarrow e^{-\beta V(r)} 
\frac{\partial V(r)}{\partial x}
\end{equation}
and the factor ${\rm{exp}}(-\beta V(r))$ so obtained excludes the unphysical 
contributions in $\delta g_{mc}({\bf{r}};\phi,\omega)$ for $r < 0$ and thus
partially compensates for the error made in the boundary condition of eq.(A.13)
as far as $\eta(\phi,\omega)$ is concerned.  Substitution of (A.14) in eq.(8) 
with the replacement (A.21) and using (A.9) leads to 
\addtocounter{eqnnum1}{1}
\begin{equation}
\eta_{mc}(\phi,\omega) = \eta_{\infty}(\phi) - \frac{2 \pi}{15} k_BT n^2
\sigma^3 f_{mc}(1;\omega)
\end{equation}
completely similar to (A.10) for $\eta(\phi,\omega)$.  Using (A.18) for 
$f_{mc}(1;\omega)$ yields the final result 
\addtocounter{eqnnum1}{1}
\begin{equation}
\eta_{mc}(\phi,\omega) = \eta_{\infty}(\phi) + \eta_0 \phi^2 \alpha_V(\omega)
\frac{1}{1 + K(\alpha)}  
\end{equation}
with $\alpha_V(\omega)$ given by (A.12), $K(\alpha)$ by (A.20) and
$\alpha = \alpha(\omega)$ by (A.8).

The result (A.23) for $\eta_{mc}(\phi,\omega)$ follows from
eq.(20) provided one uses there the low density expression for $S_{eq}
(k;\phi)$ and $\omega_H(k) = D_0k^2$.  To compare the exact expression (A.11)
for $\eta(\phi,\omega)$ with (A.23) for $\eta_{mc}(\phi,\omega)$ we 
note that for large frequencies $\omega \rightarrow \infty, 
\alpha \rightarrow \infty$ (cf.(A.8)) and $K(\infty) = 1$ (cf.(A.7), (A.19) and
(A.20)), so that then 
\addtocounter{eqnnum1}{1}
\begin{equation}
\eta_{mc}(\phi,\omega) - \eta_{\infty}(\phi) = \frac{1}{2} (\eta(\phi,\omega)
- \eta_{\infty}(\phi))
\end{equation}
For $\omega \rightarrow 0, \alpha \rightarrow 0$ (cf. (A.8)) and $K(0) = 2/3$,
so that then 
\addtocounter{eqnnum1}{1}
\begin{equation}
\eta_{mc}(\phi,\omega) - \eta_{\infty}(\phi) = \frac{3}{5} (\eta(\phi,\omega)
- \eta_{\infty}(\phi))
\end{equation}

Thus it appears that the mode coupling theory underestimates the two
particle Smoluchowski contribution to $\eta(\phi,\omega)$ by a factor 2 at
high frequencies and 5/3 at low frequencies.  The relevance of these factors
is limited in practice since for low concentrations the main contribution
to $\eta(\phi,\omega)$ comes from $\eta_{\infty}(\phi)$.
For high concentrations the factor 2 is reduced by a factor $\chi(\phi)$,
due to the replacement of $D_0$ by $D_s(\phi)$ in the two particle Smoluchowski
equation (6) (cf.section 6).

\noindent{\underline{Appendix B}}

Here we derive eq.(20) for $\eta(\phi,\omega)$ directly, using the mode-mode 
coupling approximation (mmca) for concentrated suspensions $0.3 \leq \phi \leq
0.55$, in analogy with what is done for atomic liquids$^{[19]}$. 
The basic idea behind the mmca is that fluctuations (or 'excitations')
of a given dynamical variable decay predominantly into pairs of modes
associated with conserved single-particle or collective dynamical variables
$^{[43]}$. If we restrict ourselves to the overdamped case without
hydrodynamic interactions, the only important mode is the cage diffusion mode,
i.e. the Fourier transform of the single-particle density fluctuations:
\renewcommand{\theequation}{B.\arabic{eqnnum1}}
\setcounter{eqnnum1}{1}
\begin{equation}
n({\bf k}) = \sum_{i=1}^{N} \left( e^{i{\bf k} \cdot {\bf r}_i} -
\langle e^{i{\bf k} \cdot {\bf r}_i} \rangle_{eq} \right)
\end{equation}
In this case the lowest order mmca takes into account bilinear products of
cage diffusion modes: $n({\bf k})n(-{\bf k})^{[44]}$.

We start from the Green-Kubo expression eq.(34) for $\eta(\phi,\omega)$ and
eq.(35) for the stress-stress autocorrelation function $\rho_{\eta}(t;\phi)$.
The first approximation of the mmca corresponds to the replacement of the full
evolution operator $e^{\Omega t}$ by its projection onto the subspace of the
product variables $n({\bf k})n(-{\bf k})$
\begin{equation}
\addtocounter{eqnnum1}{1}
e^{\Omega t} \approx P e^{\Omega t} P
\end{equation}
Here $\Omega$ is the N-particle Smoluchowski operator (cf. eqs.(35) and (37))
and $P$ the normalised projector operator defined by
\begin{equation}
\addtocounter{eqnnum1}{1}
P = \sum_{\bf k} \frac{ |n({\bf k})n(-{\bf k}) \rangle_{eq} \langle n({\bf k})
n(-{\bf k})| }{ 2N^2 S_{eq}^{2}(k;\phi) }
\end{equation}
where $S_{eq}(k;\phi) = \frac{1}{N} \langle n({\bf k})n(-{\bf k}) \rangle_{eq}$
is the equilibrium static structure factor and ${\bf k}$ runs over the
reciprocal lattice. From eqs.(35), (B.2) and (B.3) we find
for the stress-stress autocorrelation function
\begin{equation}
\addtocounter{eqnnum1}{1}
\rho_{\eta}(t;\phi) = \sum_{{\bf k},{\bf k}'} \frac{ \langle \Sigma_{xy}^{\eta}
n({\bf k})n(-{\bf k}) \rangle_{eq}  \langle n({\bf k})n(-{\bf k}) e^{\Omega t}
n({\bf k}')n(-{\bf k}') \rangle_{eq}  \langle n({\bf k}')n(-{\bf k}')
\Sigma_{xy}^{\eta} \rangle_{eq} }{4N^4 S_{eq}^{2}(k;\phi) S_{eq}^{2}(k';\phi)}
\end{equation}

The second approximation is to assume that the two modes appearing in the
product variables propagate independently from each other. This means that the
four-variable correlation function $\langle n({\bf k})n(-{\bf k}) e^{\Omega t}
n({\bf k}')n(-{\bf k}') \rangle_{eq}$ in eq.(B.4) can be factorised into
products of two-variable correlation functions (as already used in the
normalisation of $P$ (eq.(B.3)) giving
\begin{eqnarray}
\addtocounter{eqnnum1}{1}
\lefteqn{\langle n({\bf k})n(-{\bf k}) e^{\Omega t} n({\bf k}')n(-{\bf k}')
\rangle_{eq} = } \nonumber \\
& = & \langle n({\bf k}) e^{\Omega t} n(-{\bf k}') \rangle_{eq} \langle
n(-{\bf k}) e^{\Omega t} n({\bf k}') \rangle_{eq} + \langle n({\bf k})
e^{\Omega t} n({\bf k}') \rangle_{eq} \langle n(-{\bf k}) e^{\Omega t}
n(-{\bf k}') \rangle_{eq} = \nonumber \\
& = & N^2 F_{eq}^{2}({\bf k};t)(\delta_{{\bf k},{\bf k}'} + \delta_{{\bf k},
-{\bf k}'})
\end{eqnarray}
with $F_{eq}({\bf k};t) = \frac{1}{N} \langle n({\bf k}) e^{\Omega t}
n(-{\bf k}) \rangle_{eq}$ the equilibrium intermediate scattering function. As
outlined in section 2 the main diffusion process at long times and high
concentrations $0.3 \leq \phi \leq 0.55$ is the cage diffusion process, rather
than free diffusion. Thus the long time decay of the equilibrium intermediate
scattering function is determined by $\omega_H (k;\phi)$, the lowest eigenvalue,
given by eqs.(13) and (14), corresponding to the eigenfunction $n({\bf k})$ of a
kinetic operator defined elsewhere$^{[11-13]}$. This gives
\begin{equation}
\addtocounter{eqnnum1}{1}
F_{eq}({\bf k};t) = S_{eq}(k;\phi)e^{-\omega_H(k;\phi)t}
\end{equation}
Performing the summation over ${\bf k}'$ and changing the summation over
${\bf k}$ to an integral over
${\bf k}$ in the limit of large volume $V$, we find from eqs.(B.4)-(B.6):
\begin{equation}
\addtocounter{eqnnum1}{1}
\rho_{\eta}(t;\phi) = \frac{V}{16 \pi^3}  \int d {\bf{k}} 
[\frac{V_\eta({\bf{k}})}{S_{eq}(k,\phi)}]^2 e^{-2\omega_H(k;\phi)t}
\end{equation}
where
\begin{equation}
\addtocounter{eqnnum1}{1}
V_\eta({\bf{k}}) = \frac{1}{N} < \Sigma^{\eta}_{xy} n ({\bf{k}}) n (-{\bf{k}})
>_{eq}
\end{equation}
is the strength of the coupling between the microscopic stress tensor
$\Sigma^{\eta}_{xy}$ (eq.(36)) and two microscopic densities. To evaluate
$V({\bf{k}})$ we use that for an arbitrary function $f(r^N)$ one has: 
\begin{equation}
\addtocounter{eqnnum1}{1}
< {\bf{F}}_i f(r^N) >_{eq} = - k_BT < {\bf{\nabla}}_i f(r^N) >_{eq}
\end{equation}
where $r^N = {\bf r}_1, \cdots, {\bf r}_N$. Eq.(B.9) follows from partial
integration and using the explicit form of the
equilibrium distribution function.  Substituting eq.(36) for 
$\sum^{\eta}_{xy}$ in (B.8) and using (B.9) yields
\begin{equation}
\addtocounter{eqnnum1}{1}
V_\eta({\bf{k}}) = - \frac{k_BT}{N} \sum^N_{i=1} < r_{i,x} \frac{\partial}
{\partial r_{i,y}} n({\bf{k}}) n(-{\bf{k}}) >_{eq}
\end{equation}
From (B.1) for $n({\bf{k}})$ and the expression below (B.3) for 
$S_{eq}(k;\phi)$ follows straightforwardly
\begin{equation}
\addtocounter{eqnnum1}{1}
V_\eta({\bf{k}}) = - k_BT k_y \frac{\partial}{\partial k_x} S_{eq}(k;\phi)
\end{equation}
or equivalently,
\begin{equation}
\addtocounter{eqnnum1}{1}
V_\eta({\bf{k}}) = - k_BT \frac{k_x k_y}{k} S'_{eq}(k;\phi)
\end{equation}
Substitution in (B.7) and performing angular integrations in ${\bf{k}}$-space,
leads to the final result for $\rho_{\eta}(t;\phi)$, i.e.,
\begin{equation}
\addtocounter{eqnnum1}{1}
\rho_{\eta}(t;\phi) = \frac{(k_BT)^2 V}{60 \pi^2}  \int^{\infty}_o d k k^4
[\frac{S'_{eq}(k,\phi)}{S_{eq}(k,\phi)}]^2 e^{-2\omega_H(k;\phi)t}
\end{equation}
Then eq.(20) for $\eta(\phi,\omega)$ follows immediately from eqs.(34)
and (B.13).

\newpage

{\Large{\bf{References}}}\\

1. I. M. de Schepper, H. E. Smorenburg and E. G. D. Cohen, Phys. Rev. Lett. 
{\underline{70}}, 2178 (1993).\\
2. I. M. de Schepper and E. G. D. Cohen, Int. J. of Therm. Phys.
{\underline{15}}, 1179 (1994).\\
3. E. G. D. Cohen and I. M. de Schepper, {\em{13th Symposium on Energy 
Engineering Sciences}}, (Argonne National Laboratory (1995)).\\
4. E. G. D. Cohen and I. M. de Schepper, Phys. Rev Lett. {\underline{75}}, 
2252 (1995).\\
5. J. C. Van der Werff, C. B. de Kruif, C. Blom and J. Mellema, Phys. Rev. A 
{\underline{39}}, 795 (1989).\\
6. J. C. van der Werff and C.B. de Kruif, J. Rheol. {\underline{33}}, 421
(1989).\\
7. J. J. H. Irving and J. G. Kirkwood, J. Chem. Phys. {\underline{18}},
817 (1950).\\
8. J. O. Hirschfelder, C. F. Curtiss and R. B. Bird, {\it Molecular Theory
of Gases and Liquids}, (Wiley, (1954)), p. 652.\\
9. See, for instance, (a) D. A. McQuarrie, {\it Statistical Mechanics},
(Harper and Row, NY (1976)), p.519; (b) J. -P. Hansen and I. R. 
McDonald. {\it Theory of Simple Fluids}, (Academic Press, London (1986)), 
pp.267,268; (c) J. P. Boon and S. Yip, {\it Molecular Hydrodynamics}, 
(McGraw-Hill Inc. (1980)), p.51.\\
10. Ref.8(a) pp.250,280; ref.8(b) pp.36,95.\\
11. I. M. de Schepper, E. G. D. Cohen and M. J. Zuilhof, Phys. Lett. A 
{\underline{101}}, 399 (1984); E. G. D. Cohen, I. M. de Schepper and 
M. J. Zuilhof, Physica B {\underline{127}}, 282 (1984). \\
12. I. M. de Schepper, E. G. D. Cohen, P. N. Pusey and H. N. W. Lekkerkerker, J.
Phys. Condens. Matter {\underline{1}}, 6503 (1989); P. N. Pusey, H. N. W.
Lekkerkerker, E. G. D. Cohen and I. M. de Schepper, Physica A {\underline{164}},
12 (1990).\\
13. E. G. D. Cohen and I. M. de Schepper, J. Stat. Phys. {\underline{63}}, 241
(1991); E. G. D. Cohen and I. M. de Schepper in: {\it Recent Progress in
Many-Body Theories 3}, eds. T. L. Ainsworth, C. E. Campbell, B. E. Clements and
E. Krotscheck, (Plenum, NY (1992)), p.387.\\
14. W. B. Russel, D. A. Saville and W. R. Schowalter, {\it Colloidal
Suspensions}, (Cambr. Univ. Press (1989)), p.262-266; P. N. Pusey and R. J. A.
Tough in: {\it Dynamic Light Scattering and Velocimetry: Applications of Photon
Correlation Spectroscopy}, ed. R. Pecora (Plenum, NY (1982)); P. N. Pusey in:
{\it Liquids, Freezing and Glass Transition}, eds. J. P. Hansen, D. Levesque and
J. Zinn-Justin (North-Holland, Amsterdam (1991)), p.763.\\
15. J. K. G. Dhont, J. C. van der Werff and C. G. de Kruif, Physica A
{\underline{160}}, 195 (1989).\\
16. B. Cichocki and B. U. Felderhof, Phys. Rev. A {\underline{43}}, 5405
(1991).\\
17. {\it Handbook of Mathematical Functions}, eds. M. Abramowitz and I. A.
Stegun, (Dover Publ. Inc., NY (1972)).\\
18. Ref.8(b) p.126.\\
19. E. G. D. Cohen, Physica A {\underline{194}}, 229 (1993); E. G. D. Cohen in:
{\it 25 Years of Non-Equilibrium Statistical Mechanics}, eds. J. J. Brey, J.
Marro, J. M. Rubi, Lecture notes in Physics 445
(Springer, Berlin (1995)), p.21.\\
20. I. M. de Schepper, A. F. E. M. Haffmans and H. van Beijeren, Phys. Rev.
Lett. {\underline{57}}, 1715 (1986); T. R. Kirkpatrick, J. Non-Cryst. Solids 
{\underline{75}}, 437 (1985); T. R. Kirkpatrick and J. C. Nieuwoudt, Phys. Rev.
A {\underline{33}}, 2658 (1986).\\
21. D. Ronis, Phys. Rev. A {\underline{34}, 1472 (1986).\\
22. J. X. Zhu, D. J. Durian, J. M\"{u}ller, D. A. Weitz and D. J. Pine, 
Phys. Rev. Lett. {\underline{68}}, 2559 (1992).\\
23. B. Cichocki and B. U. Felderhof, Phys. Rev. A {\underline{46}}, 7723
(1992); B. Cichocki and B. U. Felderhof, J. Chem. Phys. {\underline{101}}, 7850
(1994).\\
24. D. Henderson and E. W. Grundke, J. Chem. Phys. {\underline{63}}, 601
(1975).\\
25. B. Cichocki and B. U. Felderhof, J. Chem. Phys. {\underline{89}, 1049 
(1988).\\
26. B. Cichocki and B. U. Felderhof, J. Chem. Phys. {\underline{89}, 3705 
(1988).\\
27. W. Hess and R. Klein, Adv. Phys. {\underline{32}}, 173 (1983).\\
28. R. Verberg, I. M. de Schepper, M. J. Feigenbaum and E. G. D. Cohen, to
be published.\\
29. E. G. D. Cohen, R. Verberg and I. M. de Schepper, to be published.\\
30. C. W. J. Beenakker, Physica A {\underline{128}}, 48 (1984).\\
31. C. W. J. Beenakker and P. Mazur, Physica A {\underline{126}}, 349 (1984).\\
32. A. Einstein, Ann. der Physik {\underline{19}}, 289 (1906);
{\underline{34}}, 591 (1911).\\
33. B. U. Felderhof, Physica A {\underline{147}}, 533 (1988).\\
34. L. D. Landau and E. M. Lifshitz, {\it Fluid Mechanics}, (Pergamon, London
(1959)), p.76.\\
35. M. L\'{o}pez de Haro, E. G. D. Cohen and J. M. Kincaid, J. Chem. Phys.
{\underline{78}}, 2746 (1983); J. M. Kincaid, M. L\'{o}pez de Haro and
E. G. D. Cohen, J. Chem. Phys. {\underline{79}}, 4509 (1983); 
H. van Beijeren and J. R. Dorfman, J. Stat. Phys. {\underline{23}}, 335 
(1980).\\
36. J. F. Brady, J. Chem. Phys. {\underline{99}}, 567 (1993).\\
37. R. J. Phillips, J. F. Brady and G. Bossis, Phys. Fluids {\underline{31}},
3462 (1988).\\
38. A. J. C. Ladd, J. Chem. Phys. {\underline{93}}, 3483 (1990).\\
39. T. N. Phung, Ph. D. thesis, California Institute of Technology (1993).\\
40. D. A. R. Jones, B. Leary and D.V. Boger, J. Coll. Int. Sci.
{\underline{147}}, 479 (1991); {\underline{150}}, 84 (1992).\\
41. Y. S. Papir and I. M. Krieger, J. Coll. Int. Sci. {\underline{34}}, 126
(1970).\\
42. B. Cichocki and B. U. Felderhof, J. Chem. Phys. {\underline{101}}, 1757 
(1994); J. F. Brady, J. Chem. Phys. {\underline{101}}, 1758 (1994).\\
43. Ref.8(b), Section 9.5.\\
44. Ref.27, Section 10.2.\\
45. W. van Megen, S. M. Underwood, R. H. Ottewill, N. St. J. Williams and
P. N. Pusey, Far. Discuss. Chem. Soc. {\underline{83}}, 47 (1987).\\
46. P. N. Pusey and W. van Megen, J. de Physique {\underline{44}}, 285 (1983).\\

\newpage
{\Large{\bf{Figure Captions}}}

\noindent 1. Reduced cage-diffusion time $\tau_c(\kappa;\phi)/\tau_P$ as a 
function of $\kappa = k \sigma$ for volume fractions $\phi$ = 0.30 (dotted
line); 0,45 (dashed line); 0.50 (solid line) and 0.55 (dash-dotted line).  For
$k \approx k^* \approx 2\pi$ the two times are of the same order of magnitude.

\noindent 2. Relative infinite frequency viscosity $\eta_{\infty} (\phi)/\eta_0$
as a function of the volume fraction $\phi$.  $\Box$ Zhu et al (ref.22);
$\times$ van der Werff et al (ref.5); $\bullet$ Cichocki and Felderhof
(ref.23) whose points were obtained by a different analyses of van der Werff 
et al's than by the authors themselves (cf. Table II).  The solid line 
corresponds to eq.(24).

\noindent 3. Relative Newtonian viscosity $\eta_N(\phi)/\eta_0$ as a 
function of the volume fraction $\phi$.  $\times$ van der Werff and de Kruif
(ref.6); $\bigtriangleup$ van der Werff et al (ref.5) (cf. Table II);
$\bullet$ Jones et al (ref.40); $\Box$ Papir and Krieger (ref.41). The solid 
line corresponds to eq.(25) and the dashed line to 
$\eta_{\infty}(\phi)/\eta_0 = \chi(\phi)$ (eq.(24)).

\noindent 4. Real (a) and imaginary part (b) of the reduced viscosity
$\eta_R^*(\phi,\omega)$ resp. $\eta_I^*(\phi,\omega)$ as a function
of $\omega \tau_1(\phi)$.  Experimental points from van der Werff et al (ref.5),
$\oplus$ for $\phi$ = 0.44, $\circ$ for $\phi$ = 0.46, $\Box$
for $\phi$ = 0.47, $\Box$ for $\phi$ = 0.48, $\bigtriangledown$ for $\phi$ =
0.51, $\star$ for $\phi$ = 0.52, $\times$ for $\phi$ = 0.54 and $\bigtriangleup$
for $\phi$ = 0.57. Theory from eqs.(20), (25) and (30). Dashed line: 
$\phi = 0.55$; solid line: $\phi = 0.50$; dotted line $\phi$ = 0.45. 
The cloud of points
in (b) near $\omega \tau_1(\phi) = 1$ should be discarded since they do not 
satisfy the Kramers-Kronig relation (ref.23).

\noindent 5. Relative real and imaginary parts of the visco-elastic viscosity,
respectively: $\eta'(\phi,\omega)/\eta_0$ ($\circ$) and
$\eta''(\phi,\omega)/\eta_0$ ($\times$), as a function of $\omega
\tau_1(\phi)$, for eight suspensions studied experimentally by van der Werff et
al (ref.5) from $\phi = 0.44$ up to $\phi = 0.57$ (cf. Table II).
In order to make a fair and realistic comparison of the theory with
experiment, keeping in mind the $4 \%$ uncertainty in the determination of
$\phi$ and the extreme sensitivity of the denominator of $\eta_{R,I}^{*}
(\phi,\omega)$ - as already pointed out by van der Werff et al$^{[5]}$ -
we assign to the experimental data an effective volume fraction $\phi^*$, 
such that $(\eta_N^{theory}(\phi^*) - \eta_{\infty}^{theory}(\phi^*)) \equiv
(\eta_N^{exp}(\phi) - \eta_{\infty}^{exp}(\phi))$, within the 
experimental uncertainty of $\phi$.
Dotted line: phenomenological results by Cichocki and Felderhof (ref.23) (only
available for $\phi$ = 0.46, 0.54 and 0.57); solid line: theory from eqs.(20)
and (29) using $\phi = \phi^*$ (cf. Table II).
                                                
\noindent 6. Ratio of $\tau_1(\phi)$ and $\tau_P$ as a function of the 
volume fraction $\phi$. Experimental points from van der Werff et al (ref.5)
(cf. Table II). Dashed line: theory from eq.(33); solid line: theory 
using eq.(41) instead of eq.(32) in eq.(30) in order to get the correct 
coefficient of the square root singularity at large frequencies (cf.section 6
and fig.7).

\noindent 7. Coefficient of the square root singularity at large frequencies
$A(\phi)$ as a function of the volume fraction $\phi$. Experimental points 
from van der Werff et al (ref.5) (cf. Table II). Dashed line: mode-mode 
coupling theory (eq.(32)); solid line: exact result starting from the
Green-Kubo relation (eq.(41)); dotted line: the theoretical result with
$D_0$ instead of $D_s(\phi) = D_0/ \chi(\phi)$ (cf.section 6)

\noindent 8. (a) Inverse relative infinite frequency viscosity 
$\eta_0/\eta_{\infty}(\phi)$ ($\bullet$ experimental points from van der 
Werff et al (ref.5); dashed line: theory from eq.(24))
and inverse relative Newtonian viscosity $\eta_0/\eta_N(\phi)$ 
($\times$ experimental points from van der Werff et al (refs.5 and 6); solid 
line: theory from eq.(25)) as a function of the volume fraction $\phi$. Dotted
line: Beenakker's expression (44c) (ref.30) (cf. Section 8, sub 3(a)).

(b) Relative short time self-diffusion coefficient
$D_s(\phi)/D_0$ as a function
of the volume fraction $\phi$. $\Box$ Zhu et al (ref.22); $\times$ Van Megen
et al (ref.45); $\bullet$ Pusey and Van Megen (ref.46). The solid line 
corresponds to eq.(45a) and the dashed line to the Beenakker and Mazur 
expression (45b) (ref.31).
             
(c) Inverse relative infinite frequency viscosity $\eta_0/\eta_{\infty}(\phi)$
($\bullet$ Zhu et al (ref.22); $\Box$ Van der Werff et al (ref.5)) and 
relative short time self-diffusion coefficient $D_s(\phi)/D_0$ ($\circ$ Zhu 
et al (ref.22); $\Box$ Van Megen et al (ref.45)) as a function of the volume 
fraction $\phi$. Solid line: theory from eq.(47b); dotted line: Beenakker 
(ref.30); dashed line: Beenakker and Mazur (ref.31).

\newpage
\begin{center}
Table I\\
Characteristic values of the model systems used$^{[5,6]}$
\end{center}

\begin{center}
\begin{tabular}{|c|c|c|c|} \hline
System & $\sigma$(nm)(DLS) & $\eta_0 (10^{12} s^{-1} m^{-2}$) & $\tau_P$ (ms)\\ 
\hline
SP 23 & 28 $\pm$ 2 & 8.68 & 0.0903\\ \hline
SSF 1 & 46 $\pm$ 2 & 5.29 & 0.400\\ \hline
SJ 18 & 76 $\pm$ 2 & 3.20 & 1.81\\ \hline
\end{tabular}
\end{center}

\vspace{10cc}

\begin{center}  
Table II\\
Parameters discussed in text.
\end{center}

\begin{center}
\begin{tabular}{|c|c|c|c|c|c|c|} \hline
$\phi$ & System & $\tau_1(\phi)/\tau_P$ & $\eta_{\infty}(\phi)/\eta_0$ & 
$\eta_N(\phi)/\eta_0$ & $A(\phi)$ & $\phi^*$\\ \hline
0.44 & SSF 1 & 0.402 & 4.99 & 12.2 & 7.69 & 0.431\\ \hline
0.46 & SP 23 & 0.421 & 5.13 & 13.1 & 8.33 & 0.438\\ \hline
0.47 & SJ 18 & 0.776 & 6.78 & 17.8 & 8.45 & 0.458\\ \hline
0.48 & SSF 1 & 0.372 & 6.36 & 17.3 & 12.1 & 0.458\\ \hline
0.51 & SJ 18 & 0.665 & 7.45 & 28.8 & 17.7 & 0.498\\ \hline
0.52 & SSF 1 & 0.834 & 7.47 & 32.7 & 18.6 & 0.508\\ \hline
0.54 & SSF 1 & 0.912 & 9.9 & 50.7 & 28.8 & 0.535\\ \hline
0.57 & SSF 1 & 3.70 & 11.5 & 139 & 44.7 & 0.593\\ \hline
0.58 & SP 23 & 3.99 & 10.0 & 187 & 60.2 & - \\ \hline
\end{tabular}
\end{center}

\newpage
Figure 1
\begin{figure}
\centerline{\psfig{figure=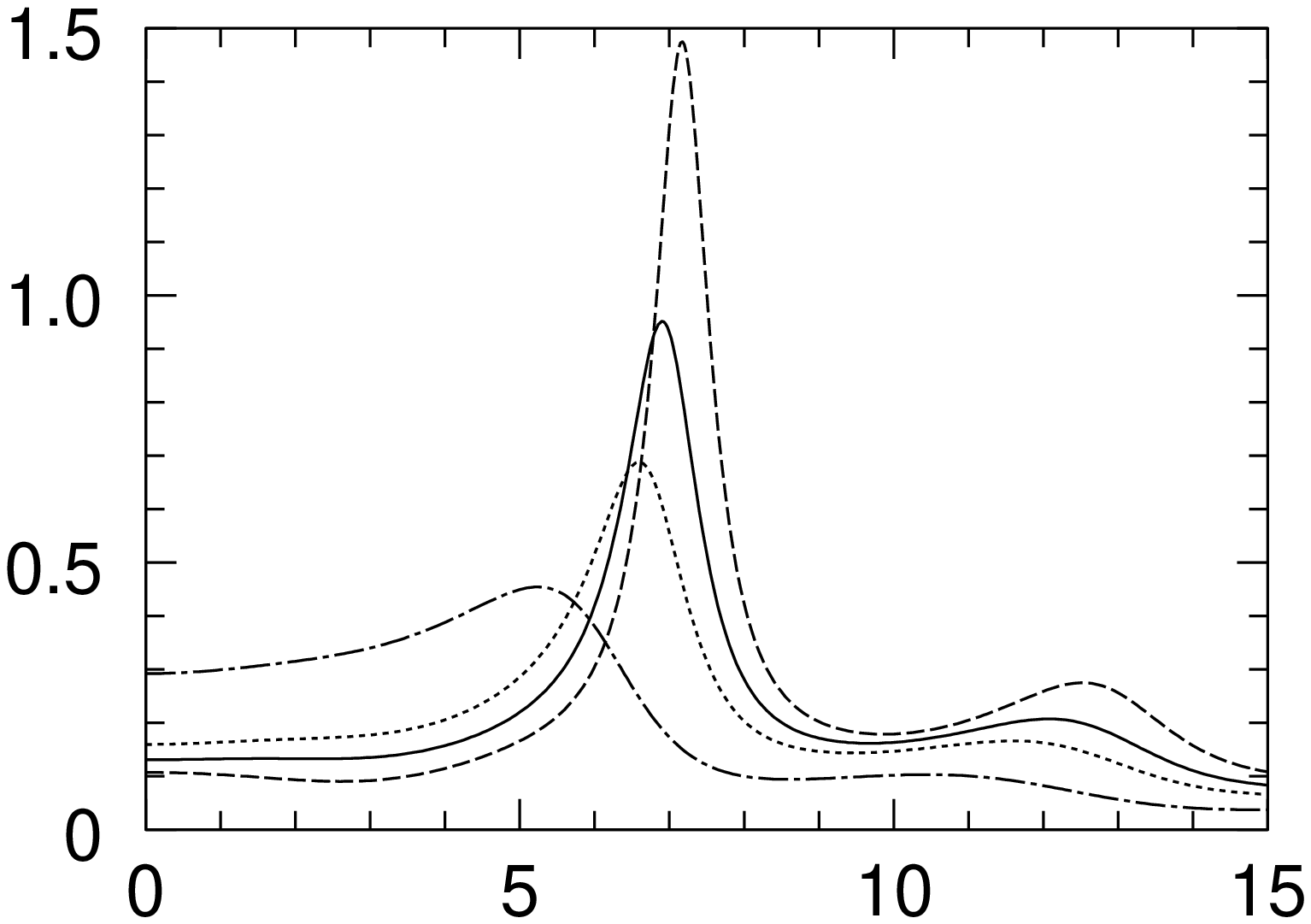}}
\end{figure}

\newpage
Figure 2
\begin{figure}
\centerline{\psfig{figure=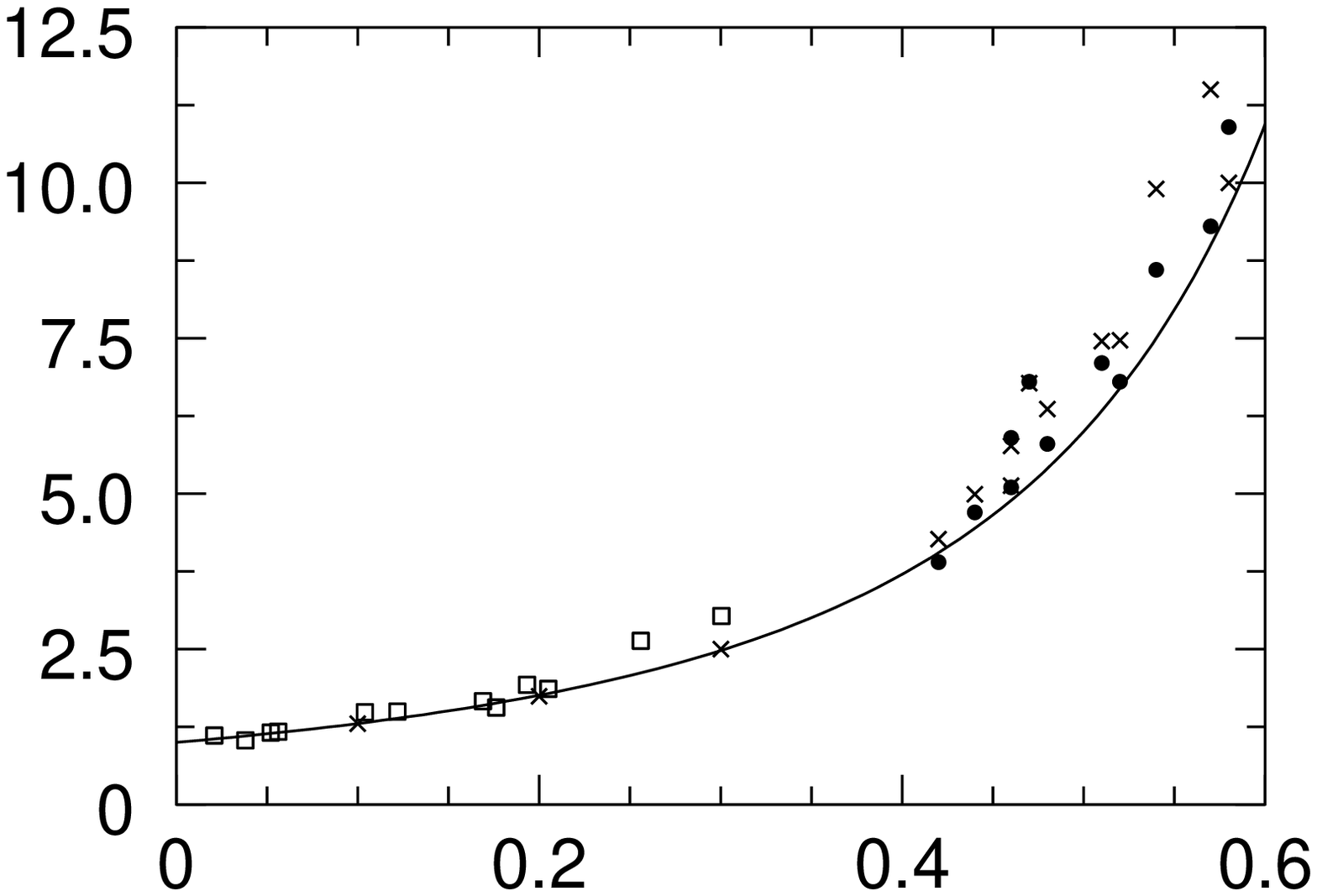}}
\end{figure}

\newpage
Figure 3
\begin{figure}
\centerline{\psfig{figure=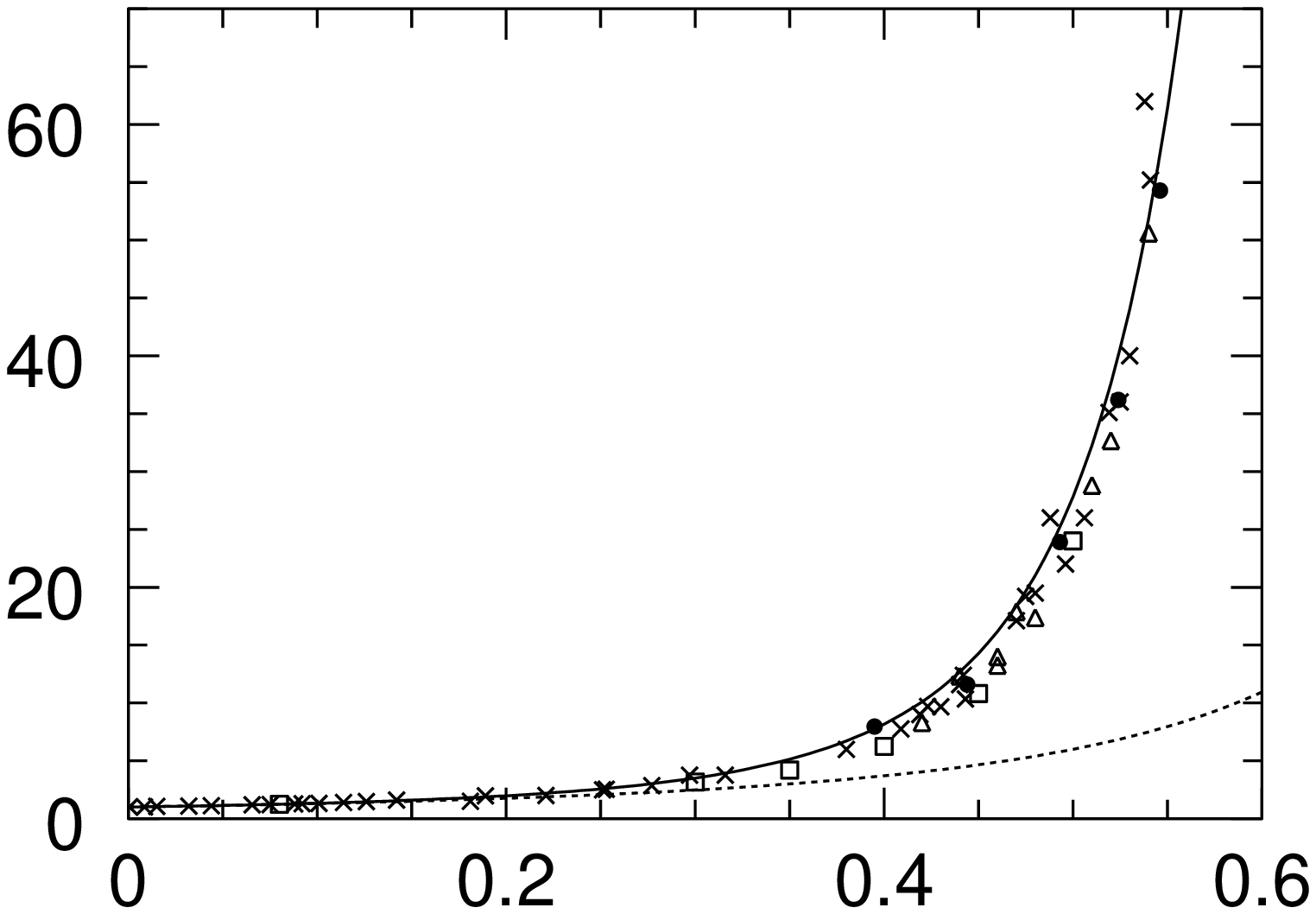}}
\end{figure}

\newpage
Figure 4a
\begin{figure}
\centerline{\psfig{figure=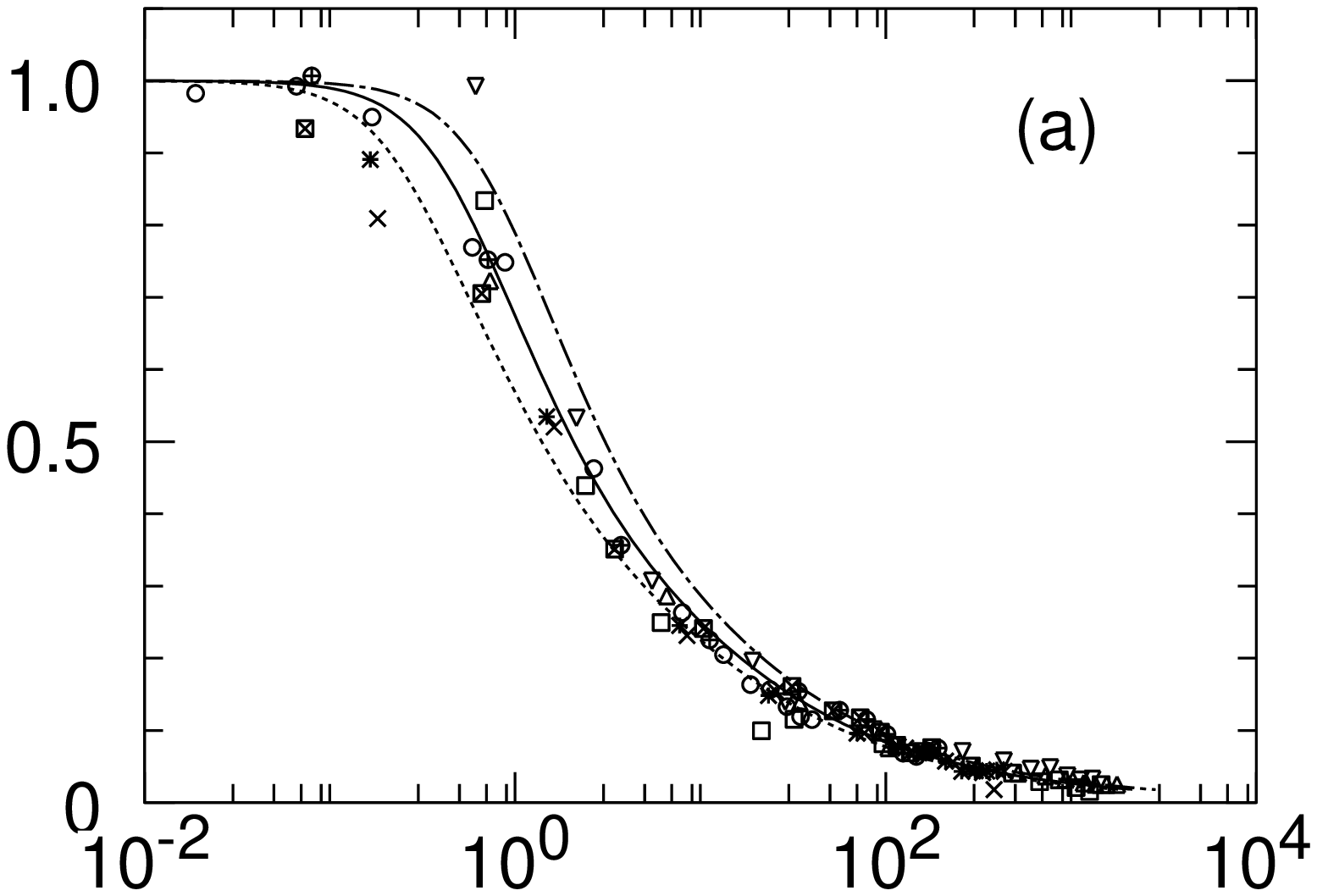}}
\end{figure}

\newpage
Figure 4b
\begin{figure}
\centerline{\psfig{figure=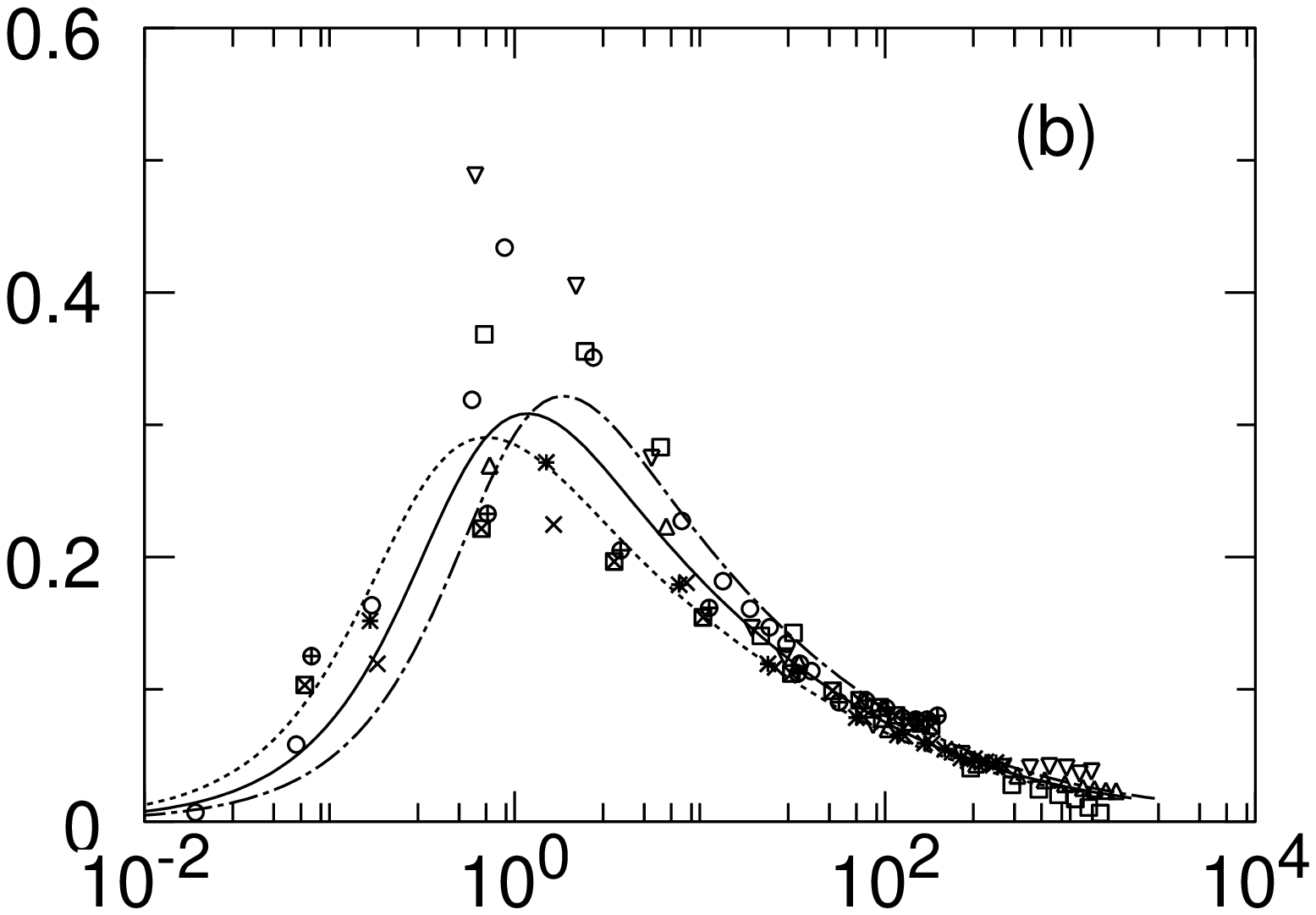}}
\end{figure}

\newpage
Figure 5a-d
\begin{figure}
\centerline{\psfig{figure=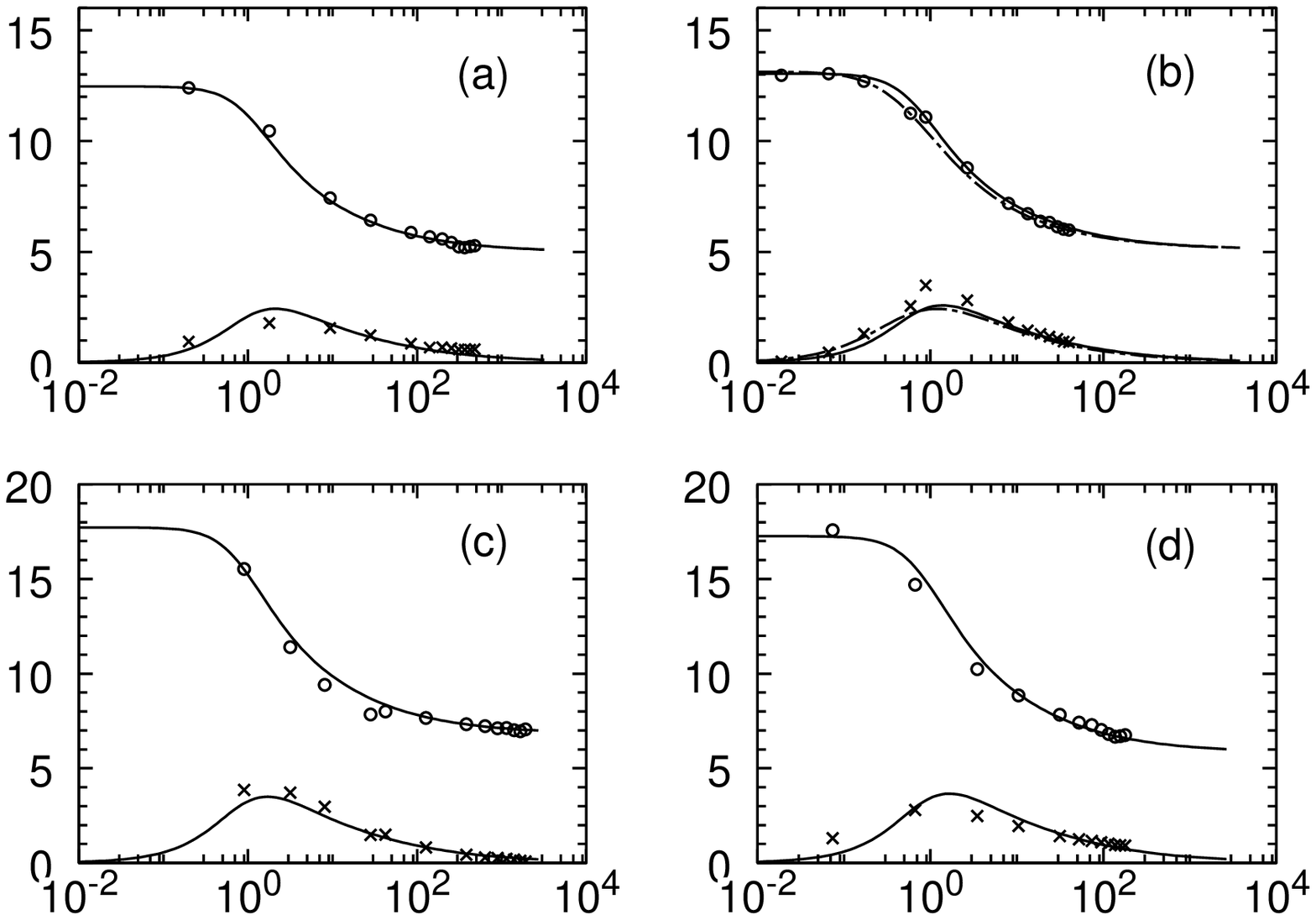}}
\end{figure}

\newpage
Figure 5e-h
\begin{figure}
\centerline{\psfig{figure=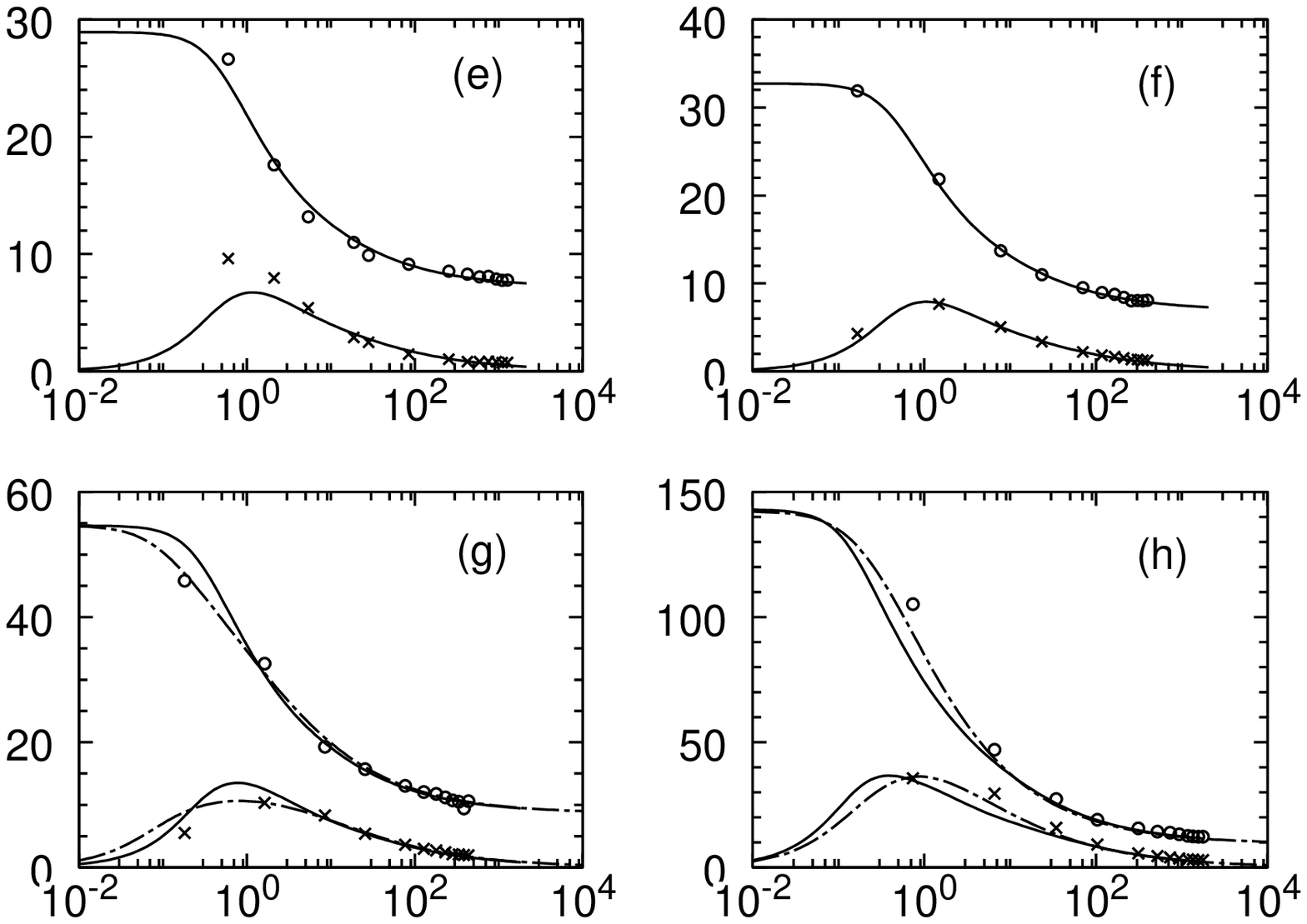}}
\end{figure}

\newpage
Figure 6
\begin{figure}
\centerline{\psfig{figure=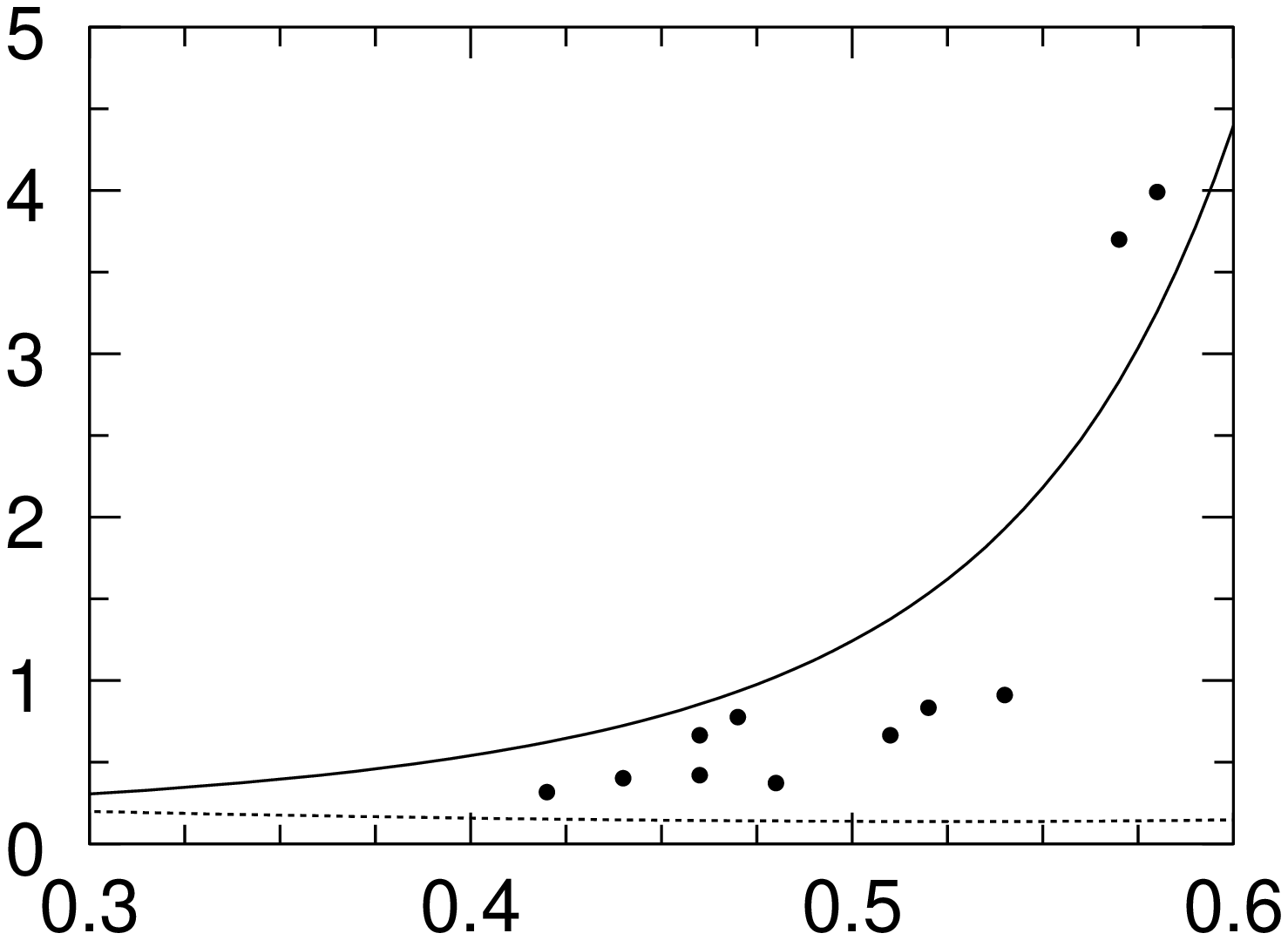}}
\end{figure}

\newpage
Figure 7
\begin{figure}
\centerline{\psfig{figure=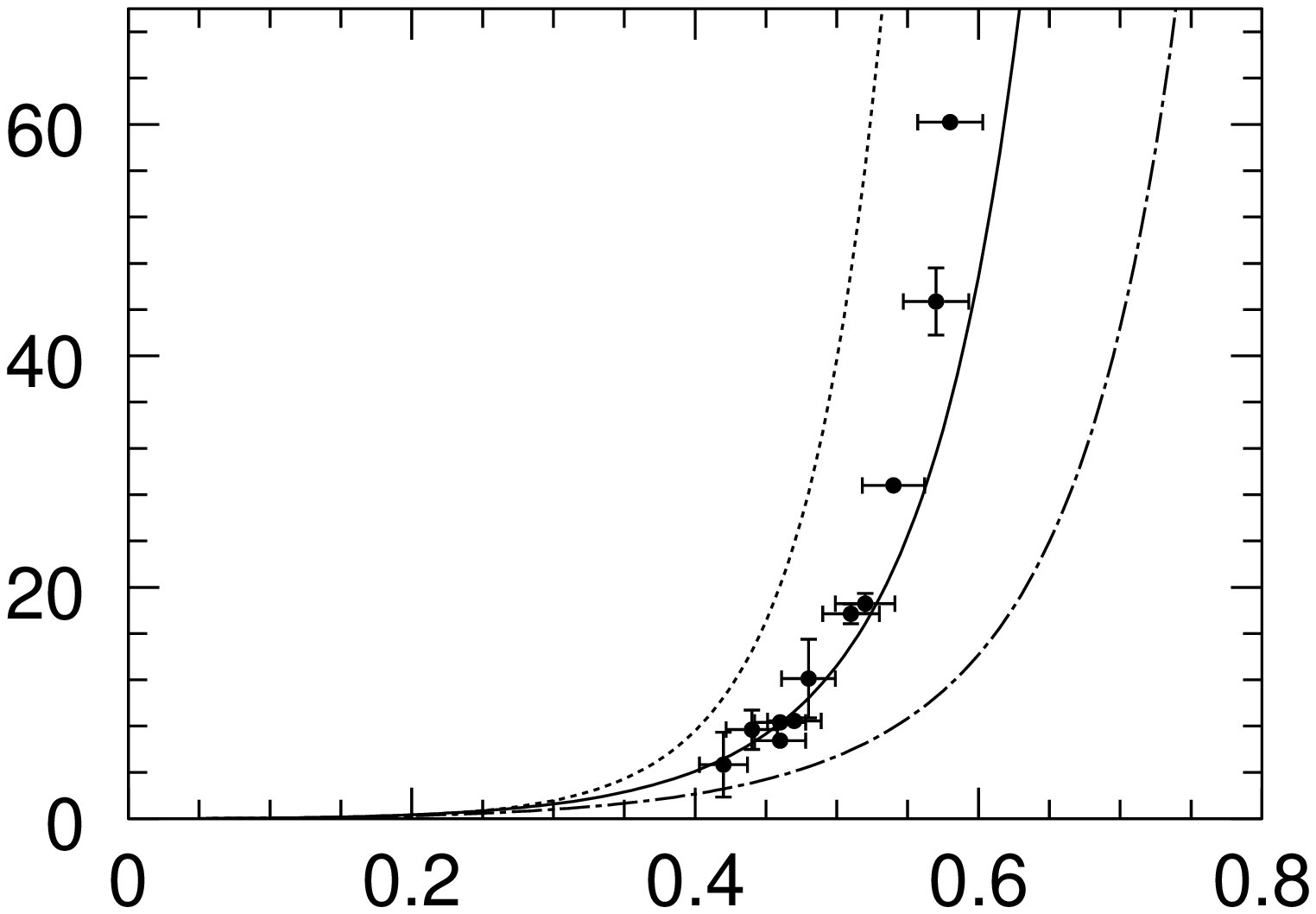}}
\end{figure}

\newpage
Figure 8a
\begin{figure}
\centerline{\psfig{figure=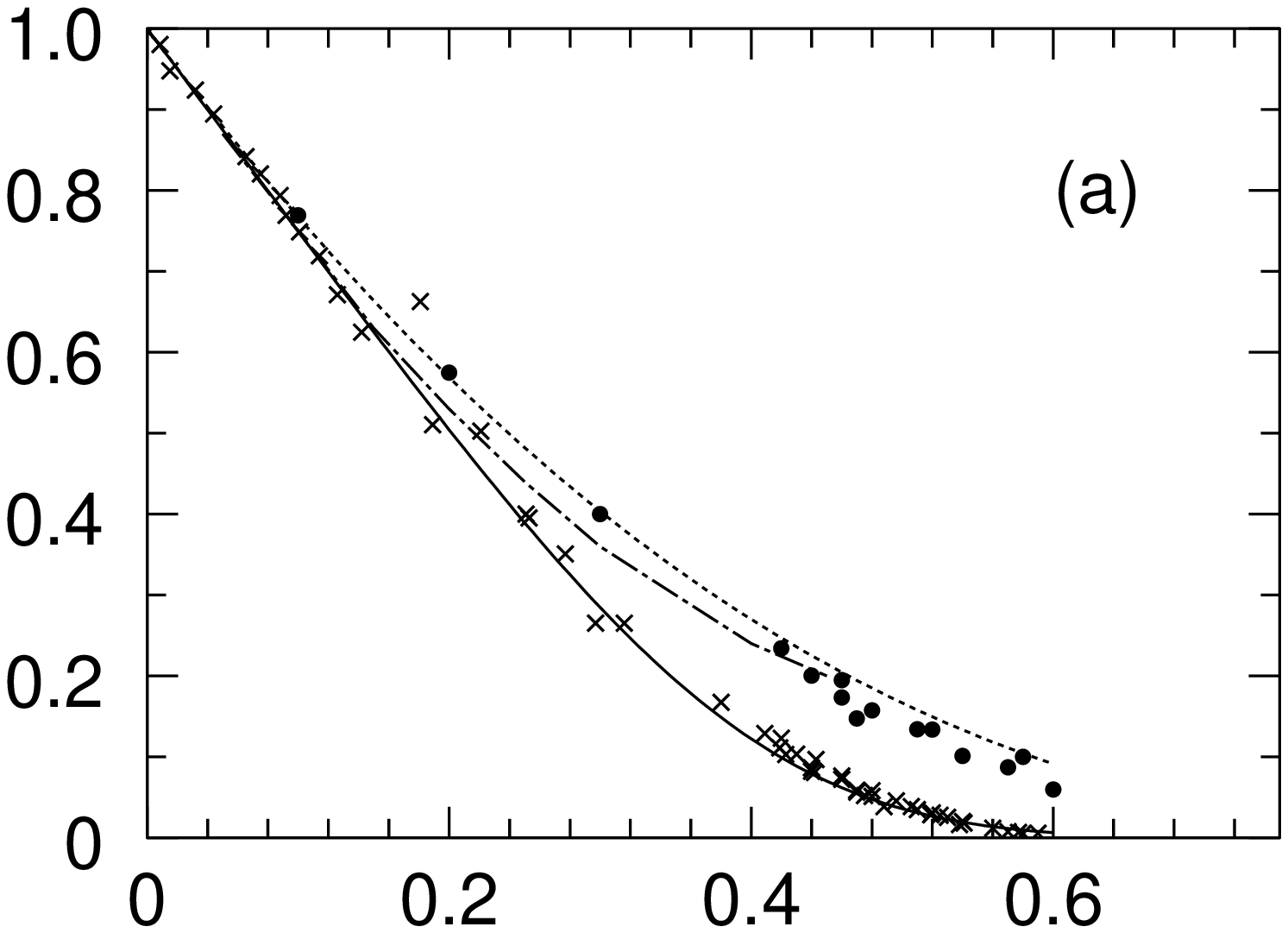}}
\end{figure}

\newpage
Figure 8b
\begin{figure}
\centerline{\psfig{figure=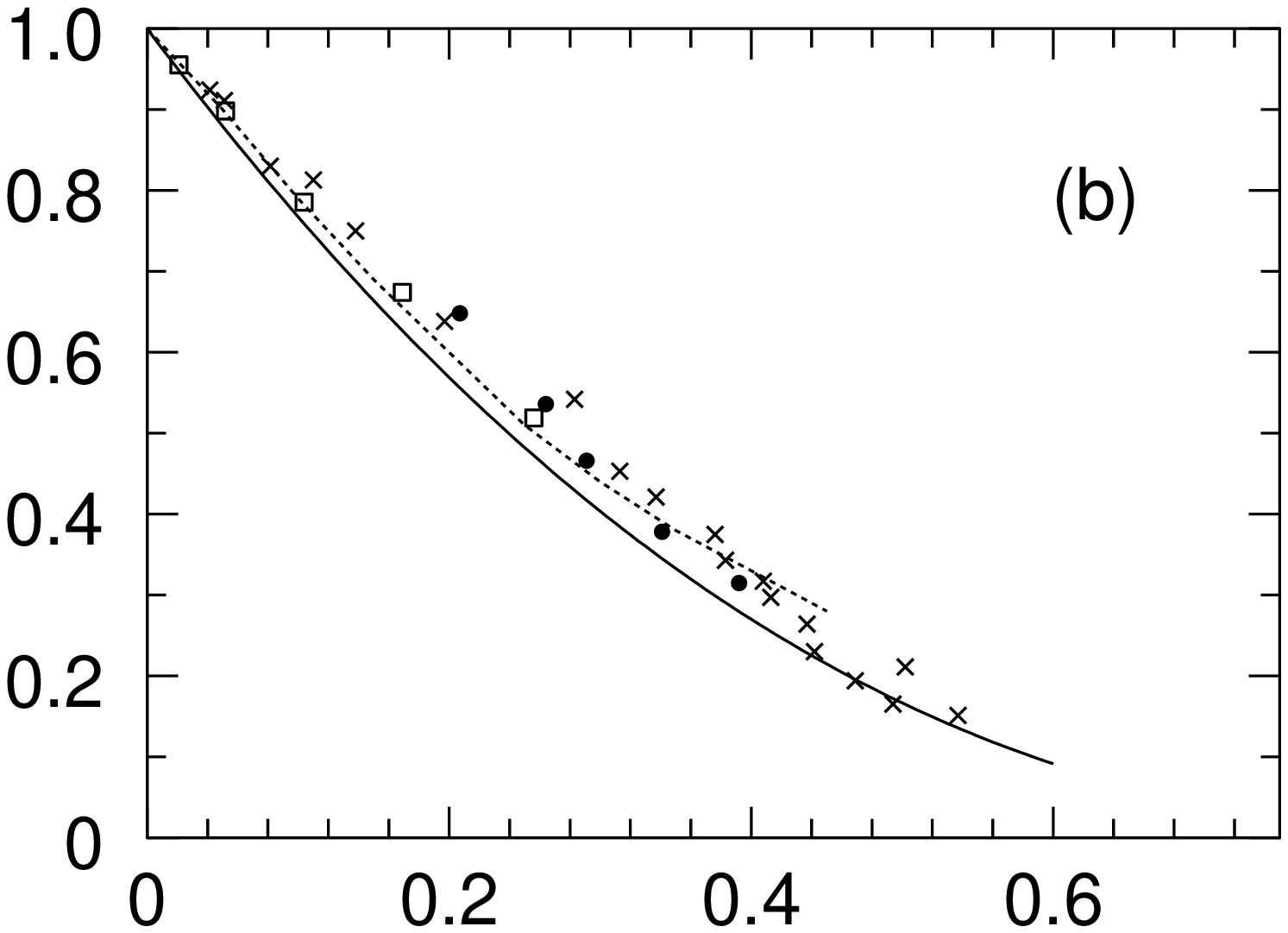}}
\end{figure}

\newpage
Figure 8c
\begin{figure}
\centerline{\psfig{figure=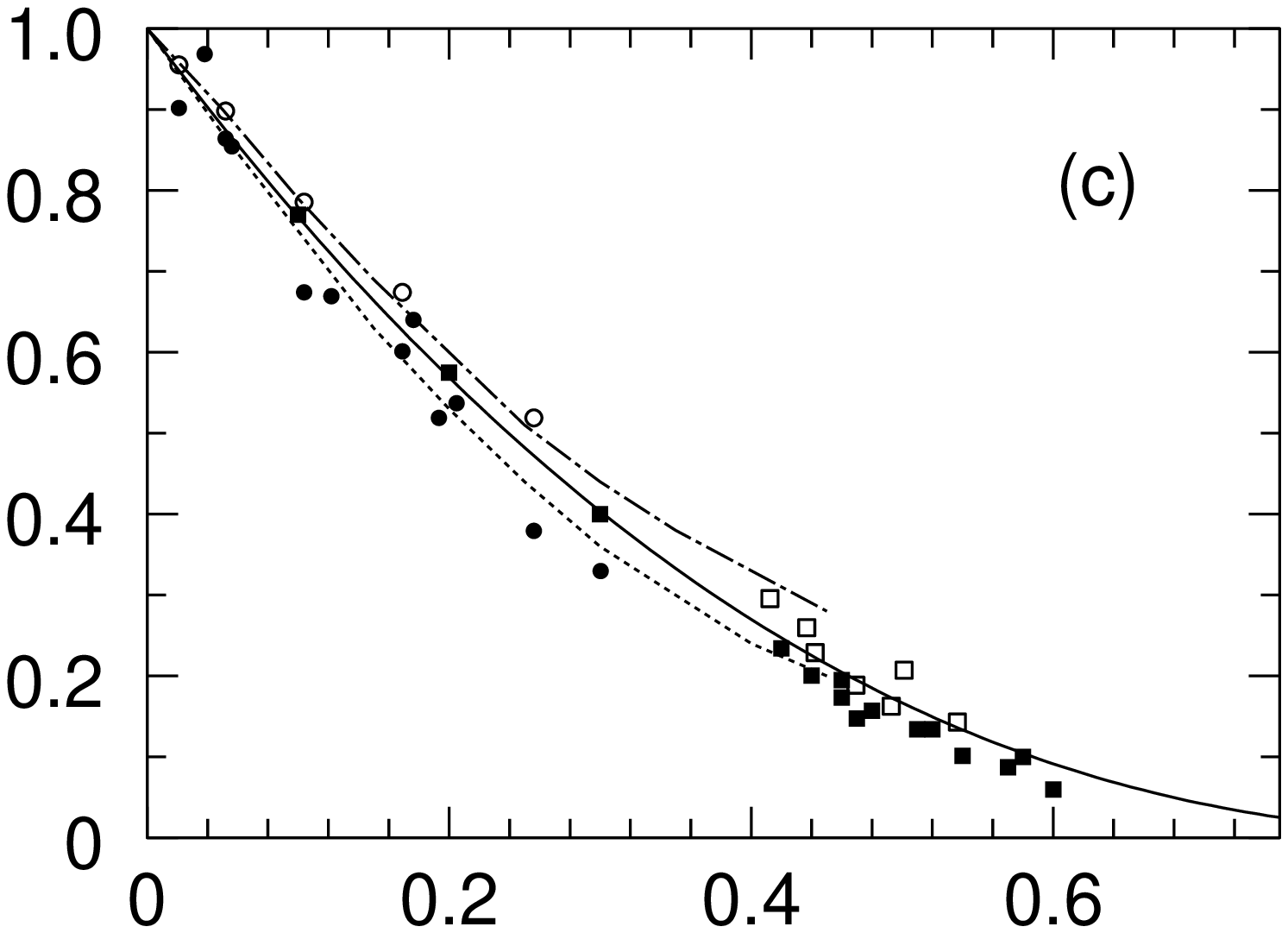}}
\end{figure}

\end{document}